\documentclass[journal=jctcce,manuscript=article]{achemso}

\usepackage[version=3]{mhchem} 
\usepackage{amsmath}
\usepackage{amsfonts,amsthm,amssymb,dsfont}
\usepackage{xr}
 \usepackage{filecontents}
\usepackage{placeins}

\newcommand{\1}{\mathds{1}}

\makeatletter
\newcommand*{\addFileDependency}[1]{
  \typeout{(#1)}
  \@addtofilelist{#1}
  \IfFileExists{#1}{}{\typeout{No file #1.}}
}
\makeatother

\newcommand*{\myexternaldocument}[1]{%
    \externaldocument{#1}%
    \addFileDependency{#1.tex}%
    \addFileDependency{#1.aux}%
}
\myexternaldocument{supp}

\author{John Strahan}
\affiliation{Department of Chemistry, University of Chicago, Chicago, IL 60637, USA}
\author{Adam Antoszewski}
\affiliation{Department of Chemistry, University of Chicago, Chicago, IL 60637, USA}
\author{Chatipat Lorpaiboon}
\affiliation{Department of Chemistry, University of Chicago, Chicago, IL 60637, USA}
\author{Bodhi P. Vani}
\affiliation{Department of Chemistry, University of Chicago, Chicago, IL 60637, USA}
\author{Jonathan Weare}
\email{weare@cims.nyu.edu}
\affiliation{Courant Institute of Mathematical Sciences,
New York University, New York, New York 10012, USA}
\author{Aaron R. Dinner}
\email{dinner@uchicago.edu}
\affiliation{Department of Chemistry, University of Chicago, Chicago, IL 60637, USA}

\title[An \textsf{achemso} demo]
  {Long-timescale predictions from short-trajectory data: A benchmark analysis of the trp-cage miniprotein}
\abbreviations{IR,NMR,UV}
\keywords{American Chemical Society, \LaTeX}

\begin{document}

\begin{tocentry}




\begin{center}
    \includegraphics[width=9cm]{figures_edited/Abstract3.png}
\end{center}

\end{tocentry}

\begin{abstract}
Elucidating physical mechanisms with statistical confidence from molecular dynamics simulations can be challenging owing to the many degrees of freedom that contribute to collective motions.  To address this issue, we recently introduced a dynamical Galerkin approximation (DGA) [Thiede {\it et al.} {\it J.\ Phys.\ Chem.} {\bf 150}, 244111 (2019)], in which chemical kinetic statistics that satisfy equations of dynamical operators are represented by a basis expansion.  Here, we reformulate this approach, clarifying (and reducing) the dependence on the choice of lag time.  We present a new projection of the reactive current onto collective variables and provide improved estimators for rates and committors. We also present simple procedures for constructing suitable smoothly varying basis functions from arbitrary molecular features.  To evaluate estimators and basis sets numerically, we generate and carefully validate a dataset of short trajectories for the unfolding and folding of the trp-cage miniprotein, a well-studied system. Our analysis demonstrates a comprehensive strategy for characterizing reaction pathways quantitatively.
\end{abstract}

\section{Introduction}


Molecular dynamics simulations enable atomic-resolution investigation of complex processes. These investigations are often carried out by direct simulation: the equations of motion are numerically integrated forward in time to generate trajectories (times series of atomic positions and, as needed, momenta) for as long as possible given available computational resources. Since most events of interest occur on timescales longer than those accessible by direct simulation, many enhanced sampling schemes have been developed to allow more extensive interrogation of an event of interest without sacrificing model fidelity. Splitting methods, for example, branch and prune a collection of simultaneously evolving trajectories to promote progress in a small number of order parameters (or collective variables, CVs) \cite{huber_weighted-ensemble_1996,dickson2010enhanced,guttenberg_steered_2012,aristoff_mathematical_2016,dinner_trajectory_2018,ffs_allen2006simulating}.  Regardless of whether trajectory data are generated by direct simulation or enhanced sampling, an essential question remains: How can these data be analyzed to yield new understanding about the process under study?



We recently introduced dynamical Galerkin approximation (DGA) \cite{thiede_galerkin_2019} to analyze trajectory data generated by direct simulation, as well as many enhanced sampling schemes.  In this approach, conditional expectations such as committor functions are cast as solutions to equations involving the operator determining the statistics of the underlying process, its transition operator.  The solution to the equation is then approximated as a linear combination of basis functions. This approach builds on an extensive literature from the last decade that shows that the eigenvalues and eigenvectors of the transition operator can be approximated from trajectory data, subject to a Markov assumption \cite{schmid_dynamic_2010,prinz2011markov,nuske_variational_2014,williams_datadriven_2015,husic_markov_2018,klus2018data,pande_everything_2010,eisner2015operator,wu_variational_2017,mardt_vampnets_2018,nuske2016variational,prinz2014spectral,delmoral2004feynman,karatzas2012brownian}.  These spectral estimation methods aim to characterize the slowest dynamical features of the system (e.g., transitions between metastable states) as eigenvectors corresponding to the largest eigenvalues of the transition operator.  When the goal is to study a particular event of interest, the indirect relationship between the eigenvectors of the generator and the event of interest is a weakness of the spectral estimation approach.  Indeed, for many complex systems the true slowest dynamical features of the system are too slow to be of any physical interest.

In contrast, the aim of DGA is \emph{not} to extract spectral information.  Instead DGA aims to compute statistics that directly characterize a particular event under study.  For example, when transitions between particular metastable states are of interest, the statistics that DGA yields can be combined within the framework of transition path theory (TPT) \cite{vanden2006transition,e_towards_2006,metzner_transition_2009} to obtain reactive fluxes and in turn reaction mechanisms. Because DGA analyzes short trajectory fragments, it can be used to process the data generated by many splitting schemes.  Alternatively, the trajectory data can be generated by seeding initial conditions for short direct simulations throughout state space.  

Though most often employed as a spectral estimation tool, Markov State Models (MSMs) have also been used to approximate TPT related quantities \cite{Noe2009,liu_markov-state_2019,meng_transition_2016}. DGA can be viewed as an extension of these MSM variants to a more general class of target quantities and to more general representations (basis set expansions) of those quantities.  However, even with parameters chosen as in MSMs, the DGA estimators introduced in this article improve upon their MSM counterparts in several ways including reduced dependence on the crucial lag time parameter and lower variance estimates of certain TPT quantities.

In our previous study \cite{thiede_galerkin_2019}, we compared diffusion map \cite{coifman_diffusion_2006} and indicator basis sets for predicting mean first-passage times and committors for the M\"uller-Brown model \cite{muller_location_1979} and the folding of the protein Fip35 using six long equilibrium trajectories from D.\ E.\ Shaw Research \cite{shaw_atomic-level_2010}.  Because there were only a few folding events within those trajectories, it was difficult to assess the performance of the method.  One goal of the present study is to generate a protein folding dataset that enables robust application of the approach and to compare different basis sets and estimators systematically.  

To this end, we study the trp-cage miniprotein, a 20-residue fast-folding artificial sequence (asn-leu-tyr-ile-glu-trp-leu-lys-asp-gly-gly-pro-ser-ser-gly-arg-pro-pro-pro-ser)
that has been studied extensively both experimentally and computationally \cite{qiu_smaller_2002,juraszek_sampling_2006,juraszek_rate_2008,marinelli2009kinetic,lindorff-larsen_how_2011,deng_how_2013,sidky_high-resolution_2019}.  In solution at 298 K, the protein folds on a 4 $\mu$s timescale and unfolds on a 12 $\mu$s timescale \cite{qiu_smaller_2002}, which makes these processes difficult but not impossible to simulate directly.  In particular, D.\ E.\ Shaw Research produced a 208 $\mu$s equilibrium simulation of the K8A mutant of trp-cage using the Anton supercomputer with the CHARMM 22* force field \cite{lindorff-larsen_how_2011}.  Although, like the Fip35 data, this trajectory contains relatively few folding events, it has been the subject of previous MSM \cite{deng_how_2013} and variational approach for Markov processes (VAMP) studies \cite{sidky_high-resolution_2019}.  These earlier studies serve as valuable points of comparison and enable us to identify CVs that provide good control over sampling.  Though DGA does not depend directly on any choice of CV, its performance is strongly affected by the quality of the dataset of sampled trajectories.  We use our chosen set of CVs together with enhanced sampling methods to generate a new dataset comprised of many short trajectories that are distributed evenly throughout the CV space.


In this article we reformulate DGA in terms of the transition operator of the underlying Markov process.  This has two primary advantages relative to our previous formulation in terms of the generator of the process \cite{thiede_galerkin_2019}.  First, it clarifies the role of lag time in DGA estimates, showing that correctly constructed estimators should have no dependence on lag time in the infinite-basis, infinite-sampling limit.  Second, the formulation in terms of the transition operator leads directly to estimators that correctly account for boundary conditions by stopping underlying trajectories appropriately.  Using our improved DGA estimators we introduce new estimators for TPT reaction rates and reactive currents.  
 To make computation of the reactive current tractable and the result readily interpretable, we introduce a projection formula for the reactive current onto a CV space which allows us to assign relative weights to transition paths in arbitrary CV spaces.  We also introduce a new procedure for constructing a basis set from arbitrary molecular features (here, primarily pairwise distances between C$_\alpha$ atoms, though we also explore CVs with delay embedding) and compare it with two basis sets that are used widely in the MSM literature:  indicator functions on molecular features and indicator functions on time-lagged independent component analysis (TICA) coordinates \cite{molgedey1994separation,perez2013identification,schwantes2013improvements}.
We show that our DGA estimators with selected basis sets can robustly yield remarkably good agreement with published results for committors and pathways, even though the total simulation time of our trp-cage dataset is only 30 $\mu$s, with a maximum trajectory length of 30 ns. The projection of the reactive currents on CVs facilitates both visualization and quantification of information about pathways, enabling immediate identification of the defining properties of transition states.   This makes our approach an efficient one for exploring mechanisms.

\section{Long time phenomena from short trajectory data}

In this section, we introduce key dynamical statistics and explain how they can be defined in terms of an evolution operator (Section \ref{secn:evolutionoperators}).  An emphasis on forms that lead directly to practical and accurate numerical estimators causes several departures from the standard presentation of this material.  We present our approach for solving the operator equations numerically by Galerkin (basis expansion) approximation \cite{thiede_galerkin_2019} (Section \ref{secn:dga}), and distinguish forward-in-time statistics (Section \ref{secn:forwardintime}) from  backward-in-time statistics involving the adjoint of the evolution operator (Section \ref{secn:backintime}); this is followed by a discussion of basis sets (Section \ref{secn:basis}) and an approach for constructing an approximately Markovian process when the molecular representation does not adequately capture the dynamics (delay embedding, Section \ref{secn:delayembedding}).  Finally, guided by TPT, we combine the dynamical statistics estimated by DGA to yield approximations of reaction rates and currents. (Section \ref{secn:tpt}).

\subsection{The transition operator and Feynman-Kac representation}\label{secn:evolutionoperators}

The dynamics of a Markov process $X(t)$ can be encoded in its associated transition operator, $\mathcal{T}^t$, 
which specifies the evolution of the expectation of a function $f$ over some interval of time $t\geq 0$:
\begin{equation}\label{eq:top}
\mathcal{T}^t f(x) = \mathbb{E}\left[f(X(t))\,|\,X(0)=x\right].
\end{equation}
The time index $t$ can be continuous or discrete.
The transition operator (also known as the Koopman operator), and in particular its eigenvectors and eigenvalues, are the key quantities in well-established methods for discovering slowly decorrelating features of a Markov process \cite{webber_error_2020}. The transition operator is also central to the DGA approach \cite{thiede_galerkin_2019}.  However, in DGA, instead of estimating the spectrum of the transition operator, the goal is to solve linear equations representing certain conditional expectations.  

In ref.\ \citenum{thiede_galerkin_2019}, we presented DGA in terms of the generator which, for a continuous time process is 
defined by the limit:
\begin{equation}\label{eq:generator}
\mathcal{L} f(x) = \lim_{t\rightarrow 0}\frac{\mathcal{T}^{t} f(x) - f(x)}{t}.
\end{equation}
For a discrete time process the limit is removed and $t$ in \eqref{eq:generator} is replaced by the unit of a single time step.
A presentation in terms of the generator has the advantage that it results in very concise equations for quantities of interest.  For example, consider the (forward) committor, $q_+(x)$, which is the probability of entering a product state $B$ before a reactant state $A$ starting from $x\notin A\cup B$:
\begin{equation}
q_+(x)={\mathbb P}[X(T_{A\cup B})\in B|x_0=x],
\end{equation}
where $T_{A\cup B} = \inf\{t> 0:\,X(t)\in A\cup B\}$ is the time of first entrance into $A\cup B$.  For $x\in A$, $q_+(x)=1$, and, for $x\in B$, $q_+(x)=0$.
The committor satisfies the Feynman-Kac relation
\begin{equation}\label{eq:genFKq+}
\mathcal{L} q_+(x) = 0\ \ \text{for}\ \ x\notin A\cup B, \qquad q_+(x)=\1_B(x)
=\begin{cases}1,& x\in B\\
    0,& x\notin B
    \end{cases}\ \ \text{for}\ \ x\in A\cup B
\end{equation}
(see Eqs.\ (18) and (19) of ref.\ \citenum{thiede_galerkin_2019}).


In this article we choose to work directly with the transition operator instead of the generator because it facilitates the implementation of numerical formulas. It also  greatly simplifies our description of TPT and clarifies the relationship between DGA and the well-established VAC approach to approximating spectral properties of the transition operator (see ref.\ \citenum{webber_error_2020}).  In the case of the committor, we integrate \eqref{eq:genFKq+} until a chosen time $\tau$ to obtain the equivalent form of the Feynman-Kac relation,
\begin{equation}\label{eq:FKq+}
    \mathcal{T}^{\tau}_{A\cup B}\,q_+(x)-q_+(x)=0
    \ \ \text{for}\ \ x\notin A\cup B, \qquad q_+(x) = \1_B(x)\ \ \text{for}\ \ x\in A\cup B.
\end{equation}
In this expression we have introduced the notation $\mathcal{T}^{t}_{A\cup B}$
for the transition operator of the stopped process $X(t\wedge T_{A\cup B})$, i.e.,  \begin{equation}\label{eq:topstop}
\mathcal{T}^{t}_{A\cup B}\, f(x) = \mathbb{E}\left[f(X(t\wedge T_{A\cup B}))\,|\,X(0)=x\right].
\end{equation}
Here and below $t\wedge T_{A\cup B} = \min\{t,T_{A\cup B}\}$, indicating that the evolution process does not proceed beyond escape. 

For a more general domain $D$ and $T_{D^\textrm{c}} = \inf\{t>0:\,X(t)\notin D\}$, the  conditional expectation 
\begin{equation}\label{eq:ce}
u(x) = \mathbb{E}\left[b(X(T_{D^\textrm{c}})) - \int_0^{T_{D^\textrm{c}}} a(X(t)) dt\,\bigg|\,X(0)=x\right]\ \ \text{for}\ \ x\in D
\end{equation}
solves the equation
\begin{equation}\label{eq:FK}
\mathcal{T}^{\tau}_{D^\textrm{c}}\,u (x) -u(x)= \int_0^\tau \mathcal{T}^{t}_{D^\textrm{c}}\, a(x) dt\ \  
    \text{for}\ \  x\in D,\qquad u(x) = b(x)\ \ \text{for}\ \  x\notin D.
\end{equation}
To obtain \eqref{eq:FKq+} for the committor, choose $D=(A\cup B)^\textrm{c}$, $b = \1_B$, and $a=0$.
In \eqref{eq:ce} and \eqref{eq:FK} we assume for simplicity that $a(x)=0$ for $x\notin D$.  For a discrete-time process the time integral in these expressions should be interpreted as a sum. 


 Crucially, \eqref{eq:FK} holds for any choice of $\tau\geq 0$ including relatively small values. For very large values of $\tau$, \eqref{eq:FK} converges to \eqref{eq:ce}.
 However, in most cases of interest, the escape time $T_{D^\textrm{c}}$ is very large, making estimation of $u$ in \eqref{eq:ce} by direct simulation of sample trajectories of $X(t)$ prohibitively expensive.  In the context of DGA, the significance of \eqref{eq:FK} is that it expresses $u$ in terms of an expectation over short trajectories.  The catch is that \eqref{eq:FK} must be ``inverted'' to solve for $u$.

\subsection{Dynamical Galerkin Approximation (DGA)}\label{secn:dga}

We now describe a Galerkin approach to approximating conditional expectations from short trajectory data.  We first introduce a ``guess'' function $\psi$ that satisfies the boundary conditions (i.e., $\psi (x)=b(x)$ for $x\notin D$).  Our approximation has the form 
\begin{equation}\label{eq:expansion}
u(x) \approx \psi(x) + \sum_{j=1}^n \phi_j(x)v_j,
\end{equation}
where $\{\phi_j(x)\}$ is a set of $n$ basis functions  satisfying $ \phi_j(x) =0$ for $x\notin D$, and $v$ is a vector of $n$ coefficients.  

\subsubsection{Forward-in-time predictions}\label{secn:forwardintime}

We begin by approximating predictions of quantities forward-in-time as in  \eqref{eq:ce} by expanding the solution $u$ of \eqref{eq:FK} at a particular user chosen value of $\tau$ called the lag time. While the solution $u$ itself is independent of $\tau$ in \eqref{eq:FK}, the quality of our approximation of $u$ with a finite basis may depend on the choice of lag time (even in the absence of sampling error).  A similar phenomenon has recently been explained in detail in the context of the VAC algorithm \cite{webber_error_2020}.
Substituting \eqref{eq:expansion} into \eqref{eq:FK}, multiplying by $\phi_i$ and integrating over the distribution of sampled points $\mu$ to form the inner product $\langle f, g\rangle = \int f(x) g(x) \mu(dx)$, 
we obtain the linear system of equations:
\begin{equation}\label{eq:exactDGA}
(C^{\tau}-C^0)v=r^{\tau},
\end{equation}
with matrices $C^{s}\in \mathbb{R}^{n\times n}$ for $s=0,\tau$,
\begin{equation}\label{eq:C}
C^{s}_{ij} = \langle \phi_i, \mathcal{T}^{s}_{D^\textrm{c}}
\, \phi_j\rangle_\mu
\end{equation}
and vector $r^{\tau}\in\mathbb{R}^n$,
\begin{equation}\label{eq:b}
r^{\tau}_i = \left\langle \phi_i, \psi(x)-\mathcal{T}^{\tau}_{D^\textrm{c}}\,\psi(x)+ \int_0^\tau \mathcal{T}^{t}_{D^\textrm{c}}\,a(x)dt \right\rangle_\mu.
\end{equation}
Given \eqref{eq:C} and \eqref{eq:b}, \eqref{eq:exactDGA} can be readily solved for $v$ by standard methods of linear algebra.  

 In models that represent molecules with high fidelity, \eqref{eq:C} and \eqref{eq:b} cannot be evaluated directly because a closed form of $\mathcal{T}^\tau_{D^\textrm{c}}$ is not known.   
 DGA overcomes this issue by approximating the action of the transition operator using short molecular dynamics trajectories: if $X(0)$ is a sample drawn from $\mu$ and $\{X(t)\}_{t=0}^{\tau}$ is a trajectory segment of length $\tau$ starting from $X(0)$, then we can estimate $C^{s}_{ij}$ (for $s=0,\tau$) and $r^{\tau}_i$ as
\begin{align}
 C^{s}_{ij} &\approx \frac{1}{M}\sum_{m=1}^M \phi_i(X^{(m)}(0))\phi_j(X^{(m)}(s\wedge T_{D^\textrm{c}})) \\
 r^{\tau}_i &\approx \frac{1}{M}\sum_{m=1}^M \phi_i(X^{(m)}(0))\left(\psi(X^{(m)}(0))-\psi(X^{(m)}(\tau\wedge T_{D^\textrm{c}})) + 
\Delta\sum_{p=0}^{N}a(X^{(m)}(p\Delta))\right)
\end{align}
where $m$ indexes trajectory segments, $\Delta$ is the sampling interval, and $N$ satisfies $N\Delta=\tau\wedge T_{D^\textrm{c}}$. To avoid overhead, it is advantageous to generate trajectories much longer than $\tau$ (but still much shorter than typical values of $T_{D^\textrm{c}}$) and use a rolling window to generate short trajectories of length $\tau$.  We further note that in practice configurations are not saved at every molecular dynamics step.  This limits the resolution of both the lag time and the stopping time, which we take to be the time of the first saved configuration outside the domain $D$.

\subsubsection{Adjoints, the steady state, and backward-in-time predictions}\label{secn:backintime}

To compute many important quantities we need not only to solve equations involving the transition operator but also equations involving its adjoint $(\mathcal{T}^t)^\dagger_\mu$ in the $\mu$-weighted inner product, which by definition satisfies 
\begin{equation}\label{eq:adjoint}
    \int f(x)\mathcal{T}^tg(x)\,\mu(dx) = \int g(x)(\mathcal{T}^t)^\dagger_\mu f(x)\, \mu(dx).
\end{equation}

One such equation is for the change of measure $w= d\pi/d\mu$, which can be used to reweight from the sampling distribution $\mu$ to the stationary distribution $\pi$:
\begin{equation}\label{eq:reweight}
\int f(x)\pi(dx) = \int f(x) w(x) \mu(dx),
\end{equation}
assuming $\mu$ and $w$ are normalized such that $\int w(x)\mu(dx)=1$.
Owing to the time translational invariance of averages over the stationary distribution $\pi$, \eqref{eq:adjoint}, and \eqref{eq:reweight},
the change of measure satisfies the equation
\begin{equation}\label{eq:changeofmeasure}
\begin{split}
(\mathcal{T}^\tau)^\dagger_\mu w(x) - w(x)=0.
\end{split}
\end{equation}
\eqref{eq:changeofmeasure} can be solved analogously to \eqref{eq:FK}, but, in this case, there are no boundary conditions. The introduction of a basis leads to a linear system of equations  of the form
\begin{equation}
    (\bar{C}^\tau - \bar{C}^0)^\top v=0,
\end{equation}
with $\bar{C}^s$ (for $s=0,\tau$) differing from $C^s$ only in the choice of basis (which is no longer restricted to $D$) and the use of $\mathcal{T}^s$ in place of $\mathcal{T}^s_{D^\textrm{c}}$; $\top$ denotes the transpose.  We note that by including $\phi_1(x)=1$ in the basis we can guarantee that the equation for $v$ has a solution.  
Given an approximate $w$, \eqref{eq:reweight} can be computed as
\begin{equation}
\int f(x)\pi(dx)
\approx \sum_{j=1}^M f(X^{(j)}(0))  w(X^{(j)}(0)),
\end{equation}
with the weights normalized such that
\begin{equation}
    \sum_{j=1}^M w(X^{(j)}(0))=1.
\end{equation}
That the change of measure can be estimated from short nonequilibrium trajectory data was previously observed in ref.\ \citenum{wu_variational_2017}.

Another important quantity expressible in terms of an equation involving an adjoint of the transition operator is the backwards committor
\begin{equation}
q_-(x) = \mathbb{P}\left[X(-T^-_{A\cup B})\in A\,|\,X(0)=x\right],\quad T^-_{A\cup B} = \inf\{t> 0:\,X(-t)\in A\cup B\},
\end{equation}
where  $X(-t)$, $t\geq 0$ is the steady-state backward-in-time process governed by the transition operator
\begin{equation}\label{eqn:piadjoint}
\mathcal{T}^{-t}f(x) = (\mathcal{T}^t)^\dagger_\pi f(x)= \frac{1}{w(x)}(\mathcal{T}^t)^\dagger_\mu[fw](x)
\end{equation}
(the last equality can be verified using \eqref{eq:adjoint}).
The backward committor is the probability that a trajectory currently at position $x$ last came from the reactant state $A$ rather than the product state $B$. 
It satisfies the Feynman-Kac relation
\begin{equation}\label{eqn::qback}
\mathcal{T}^{-\tau}_{A\cup B}\, q_-(x)-q_-(x)=0\ \ \text{for}\ \ x\notin A\cup B,
\qquad  q_-(x) = \1_A(x)\ \ \text{for}\ \ x\in A\cup B.
\end{equation}
Consistent with our definition of $\mathcal{T}^{t}_{A\cup B}$ above, $\mathcal{T}^{-t}_{A\cup B}$ is the transition operator for the steady-state backward-in-time process stopped upon first entrance in $A\cup B$.

To expand and approximate $q_-$ according to the DGA recipe described above, we need to estimate $\mu$-weighted inner products involving $\mathcal{T}^{-t}_{A\cup B}$.
To that end we note that, as long as $g=0$ on  $A\cup B$,
\begin{equation}\label{eq:backip}
\left\langle g, \mathcal{T}^{-t}_{A\cup B}\,f\right\rangle_\mu =
\int \mathbb{E}\left[f(X(S_{A\cup B}(t)))\frac{g(X(t))}{w(X(t))}
\,\bigg{|}\,X(0)=x\right]w(x) \mu(dx)
\end{equation}
where
\begin{equation}
\label{eqn:saub}
S_{A\cup B}(t) = \sup\{s\leq t: \, X(s)\in A\cup B\}
\end{equation}
(with $S_{A\cup B}(t)=0$ if $X(s)\notin A\cup B$ for all $0\leq s\leq t$).
We provide a derivation of \eqref{eq:backip} in Appendix \ref{secn:innerproductderivation}.  Just as for the forward committor, we expect that use of a sampling measure $\mu$ with high resolution in transition regions will lead to higher approximation accuracy (i.e., better ability of a finite basis to capture the dynamics).  However, in our experience the factor of $w^{-1}(X(t))$ in \eqref{eq:backip} leads to  significant sampling errors for larger values of $t$.  For our backward committor calculation we therefore weight inner products by $\pi$, using the formula
\begin{equation}\label{eq:backip2}
\left\langle g, \mathcal{T}^{-t}_{A\cup B}\,f\right\rangle_\pi =
\int \mathbb{E}\left[f(X(S_{A\cup B}(t))){g(X(t))}
\,\bigg{|}\,X(0)=x\right]w(x) \mu(dx).
\end{equation}
 \eqref{eq:backip2} allows inner products involving $\mathcal{T}^{-t}_{A\cup B}$ to be computed using forward trajectories of $X$ initiated according to $\mu,$ i.e., exactly the same ingredients required to make forward-in-time predictions by DGA.


Following our procedure for forward quantities outlined in Section \ref{secn:dga}, given a guess function $\psi$ satisfying $\psi=1$ on $A$ and $\psi=0$ on $B$ and basis functions $\phi_j$ that are zero on $A\cup B$, we can build an approximation
\begin{equation}
q_-(x) \approx \psi(x) + \sum_{j=1}^n \phi_j(x)v_j
\end{equation}
by solving 
\begin{equation}
    (C^{-\tau} - C^0) v= r^{-\tau}
\end{equation}
with
\begin{equation}
C^{-\tau}_{ij} = \left\langle\phi_i, \mathcal{T}^{-\tau}_{A\cup B}\,\phi_j \right\rangle_\pi= \int \mathbb{E}\left[\phi_j(X(S_{A\cup B}(\tau))){\phi_i(X(\tau))}
\,\bigg{|}\,X(0)=x\right]w(x) \mu(dx)
\end{equation}
and
\begin{align}
r^{-\tau}_{i}
&=\left\langle \phi_i,\, \psi \right\rangle_\pi- \left\langle \phi_i,\, \mathcal{T}^{-\tau}_{A\cup B}\,\psi \right\rangle_\pi \nonumber\\
&=  \int \mathbb{E}\left[\left(\psi(X(\tau))-\psi(X(S_{A\cup B}(\tau)))\right){\phi_i(X(\tau))}
\,\bigg{|}\,X(0)=x\right]w(x) \mu(dx)
\end{align}
where the second equality in each display follows from \eqref{eq:backip2}.


Along with the forward committor $q_+$ and the stationary change of measure $w$, the backward committor is a key ingredient of TPT.  In Section \ref{secn:tpt} we  describe how DGA estimates of these quantities can be combined with TPT to reveal key properties of steady-state transition paths from the reactant state $A$ to the product state $B$.  However, before that, we complete our presentation of DGA with a discussion of molecular representations and basis sets, with emphasis on those that we employ in the present study to analyze trp-cage miniprotein unfolding and folding.

\subsubsection{Basis functions}
\label{secn:basis}
A key determinant of the performance of DGA is the choice of basis set. Constructing a basis set that respects the boundary conditions of the problem and captures the dynamics with relatively few functions requires care.  Here we discuss how we generated the basis sets that we compare later in our numerical experiments, and explain why we chose them over alternatives.

In addition to the choice of functions, there is also a choice of molecular representation (i.e., the features that serve as inputs to the functions).  Although molecular dynamics trajectories are generally recorded as sequences of Cartesian coordinates, the inputs to the basis functions are generally internal coordinates.  This removes the effects of trivial translations and rotations, and it can improve the statistics.  The internal coordinates that we use are pairwise distances between C$_\alpha$ atoms greater than two sequence positions apart; for trp-cage, there are 153 such distances.  In other words, the process $X(t)$ to which we apply DGA (and TPT) is the length 153 vector of pairwise distance values.  In our tests we found that including additional features, such as backbone dihedral angles, did not improve performance.
 We assume that the reactant state $A$ and product state $B$ of interest can be characterized in terms of these variables.  We construct basis functions of these variables that satisfy the homogeneous boundary condition on the domain  $D=(A\cup B)^\textrm{c}$.


In this work, we compare three choices of basis set: indicator functions on the pairwise distances, indicator functions constructed on the top 10 TICA coordinates\cite{molgedey1994separation,perez2013identification,schwantes2013improvements} computed from the pairwise distances at a lag time of 0.5 ns, and smooth functions of pairwise distances that satisfy the boundary conditions.  We refer to these henceforth as the distance indicator, TICA indicator, and modified distance basis sets. 
We constructed the distance indicator and TICA indicator basis sets and their guess functions as follows:
\begin{enumerate}
    
    \item For the distance indicator basis set, we constructed 200 indicator functions by mini-batch $k$-means clustering as implemented in PYEMMA on the values of the 153 pairwise distances.  For the TICA indicator basis set, the clustering was performed on the top 10 TICA coordinates constructed on the pairwise distances.
    
    \item We retained all resulting indicator functions with non-zero regions fully contained in $(A\cup B)^c$ as the basis set.  We split any indicator functions with non-zero regions overlapping with $A$ or $B$, and we redefined them to be non-zero only in the portions in $(A\cup B)^c$.  For the change of measure calculations, boundary conditions are not present, so we used all indicator functions unmodified.
    
    \item For the forward committor calculation we took the the guess function to be $\psi(x)= \1_B(x)$.  For the backward committor calculation we took the guess function to be $\psi(x) = \1_A(x)$.
\end{enumerate}
With an indicator basis, the DGA and MSM estimator (with appropriate state definitions) of the forward committor $q_+$ and change of measure $w$ become similar \cite{thiede_galerkin_2019}.  We note however that the DGA (as formulated here) and MSM approaches diverge both in DGA's use of stopped trajectories and in the way $q_+$ and $w$ (and $q_-$) are used to estimate TPT quantities as described in Section \ref{secn:tpt}.

We constructed the distance basis set and its guess function as follows:
\begin{enumerate}
    
    \item We computed $d_A$ and $d_B$ as the distance in feature space (i.e. in {153}-dimensional Euclidean space) to the sampled points in states $A$ and $B$, respectively.
    
    \item We set $h(x)=d_Ad_B/(d_A+d_B)^2$, which obeys the homogeneous boundary conditions by construction.
    
    \item We computed basis functions obeying the boundary conditions by multiplying each coordinate of the pairwise distance vector $x$ by $h(x)$: $\phi_i(x)=x_i\, h(x)$.  For the change of measure calculation, we use $\phi_i=x_i $ and add the constant function into the set of chosen features. 
    
    \item To remove any linear dependencies introduced by enforcing the boundary conditions, and to ensure numerical stability, we orthogonalized the basis set ${\phi_i}$ with respect to the sampling measure (up to sampling error) using a singular value decomposition. 
    
    \item For the forward committor calculation we took the guess function to be $\psi(x)=d_B^2/(d_A+d_B)^2$. For the backward committor calculation we took the guess function to be $\psi(x)=d_A^2/(d_A+d_B)^2$.
\end{enumerate}
Although here we use the backbone pairwise distances, we note that this construction procedure could be used to generate basis sets obeying the homogeneous boundary conditions for a choice of variables other than the pairwise distances such as dihedral angles, radial basis functions, or soft indicator functions.

The indicator and TICA basis sets are the most widely used in the MSM literature.  Various alternatives have been proposed specifically in the context of spectral estimation \cite{schutte2011markov,vitalini2015basis,boninsegna2015investigating,weber2017set}.  In our previous work \cite{thiede_galerkin_2019}, we considered a basis set based on diffusion maps \cite{coifman_diffusion_2006}.  Due to the size of our trp-cage dataset ($\sim$10$^6$ datapoints), the $O(N^3)$ scaling of the matrix diagonalization associated with the diffusion map proved prohibitively computationally costly without subsampling and out of sample extension.   

\subsubsection{Delay Embedding}
\label{secn:delayembedding}
Application of DGA as described so far assumes that the underlying process $X(t)$ is Markovian; the conditional expectations that DGA seeks to approximate are not fully defined if $X(t)$ is not Markovian.  Yet, in the previous section we described an approach to building a basis set for DGA consisting of functions of only a subset of the full collection of variables  (selected pairwise distances).  Though the dynamics of this subset are not strictly Markovian, in Section \ref{secn:trpcage} we show that, at least in the specific context of the trp-cage system, the remaining degrees of freedom relax sufficiently fast that DGA yields accurate results.

However, in some circumstances, one may only have access to a small number of variables that are insufficient to specify the dynamics.  This situation is typical when the data are from an experiment.  In this case, we can construct a more expressive representation of the system from time-lagged images, i.e., if $X(t)$ is not itself Markovian we can instead apply DGA to the augmented process $\bar{X}(t) = (X(t-M\delta),X(t-(M+1)\delta),\dots, X(t))$ \cite{thiede_galerkin_2019}.  For large enough $M$ one can expect $\bar{X}(t)$ to be nearly Markovian.  

In Section \ref{secn:delaydemo},
 we show that delay embedding can significantly improve DGA estimates when a small number of CVs is used to characterize molecular configurations.  Writing the values of the CVs at time $t$ as the vector $X(t)$, we construct the delay embedded process $\bar{X}(t)$.  We then construct a basis set following the recipe in Section \ref{secn:basis} for the modified distance basis, but replacing $X$ with $\bar X$.  We then extend other functions $f$ of the CV space  to the delay-embedded space by $f(\bar{X}(t))=f(X(t-\lfloor M/2\rfloor\delta))$.  This allows us to extend the states $A$ and $B$ (which can both be defined in terms of the CVs) as well as the functions $a$ and $b$ in \eqref{eq:ce}. We then apply DGA as outlined above directly on the delay-embedded space.

\subsection{Reaction rates and currents}\label{secn:tpt}
Estimates of rates from simulations are frequently of interest because they can be compared directly with experimental measurements, and they can provide indirect information about mechanisms.  TPT in principle provides not just rate estimates but reactive currents or fluxes, which provide direct information about mechanisms.  However, previous calculations of reactive current have been limited to toy models and depictions of the reactive flux between metastable states can been difficult to interpret. Working within the TPT framework and building upon DGA approximations of $w$, $q_+$, and $q_-$, in this section we introduce robust estimates of the reaction rate and of an easily interpretable projection of the reactive current onto CVs (as opposed to over the network of metastable states).

There are various expressions for the rate in TPT.  One approach is based on the rate at which trajectories transition from $A$ to $B$, $R_{AB}$.  
If $U$ is any set for which $A\subset U$ and $B\subset U^\textrm{c}$ then, for a continuous time process,
 \begin{align}\label{eqn:bareRate_S}
     R_{AB}&=\lim_{t\rightarrow 0}\frac{1}{t}\int \left(\1_U(x)\mathcal{T}^{t}[\1_{U^\textrm{c}}q_+](x)-\1_{U^\textrm{c}}(x)\mathcal{T}^{t}[\1_{U}q_+](x)\right) q_-(x)\pi(dx) \nonumber \\
     & = \lim_{t\rightarrow 0}\frac{1}{t}\int \left(\1_U(x)\mathcal{T}^{t}q_+(x)-\mathcal{T}^{t}[\1_{U}q_+](x)\right) q_-(x)\pi(dx),
 \end{align}
where the second line is obtained by noting that $\1_{U^\textrm{c}}(x) = 1 - \1_{U}(x)$.
Here and below, for a discrete time process the limit is removed and $t$ is replaced by the unit of a single time step.  

Expression \eqref{eqn:bareRate_S} simply counts trajectories with forward crossings of the surface dividing $U$ and $U^\textrm{c}$, weighted by their probabilities that they start in $A$ and end in $B$.  Consequently, when using this formula to estimate rates from data, only those trajectories that cross the surface dividing $U$ and $U^\textrm{c}$ contribute.  Because these trajectories are generally a small fraction of the data, this results in relatively large variances in estimates.  We can obtain considerably better estimates by considering the isocommittor surfaces: $U(z)=\{x:q_+(x)\leq z, \ x \in D\}$ for $z\in(0,1)$, and noting that $R_{AB}$ is independent of $z$.  Integrating \eqref{eqn:bareRate_S} with respect to $z$ and using linearity of $\mathcal{T}^t$ yields
\begin{align}\label{eq:Nint}
R_{AB} &= \int_0^1R_{AB}\,dz \notag \\ &= \lim_{t\to0}\frac1t\int\bigg\{\mathcal{T}^tq_+(x)\int_0^1\1_{[0,z]}\big(q_+(x)\big)\,dz \nonumber -\mathcal{T}^t\left [q_+ \int_0^1\1_{[0,z]}(q_+)\,dz\right ](x)\bigg\}q_-(x)\pi(dx) \nonumber\\
&=\lim_{t\to0}\frac1t\int\bigg\{\mathcal{T}^tq_+(x)\big[1-q_+(x)\big] \nonumber - \mathcal{T}^t\big[q_+(1-q_+)\big](x)\bigg\}q_-(x)\pi(dx) \nonumber\\
&= \lim_{t\rightarrow 0} \frac{1}{t}\int \left(\mathcal{T}^{t}q_+^2(x)-q_+(x)\mathcal{T}^{t}q_+(x)\right) q_-(x)\pi(dx),
\end{align}
where we have made use of the fact that the integral of the Heaviside function (which enters through the indicator functions) is the ramp function.
This expression for $R_{AB}$ immediately suggests the estimator:
\begin{equation}\label{Rate_estimator}
    R_{AB} \approx\frac{1}{t}\sum_i q_+(X^{(i)}(t\wedge T_{A\cup B}))\left[q_+(X^{(i)}(t\wedge T_{A\cup B}))-q_+(X^{(i)}(0))\right]q_-(X^{(i)}(0))w(X^{(i)}(0))
\end{equation}
for some small choice of $t$. Note the use of stopped trajectories in \eqref{Rate_estimator}.  For very small values of $t$ the inclusion of the stopping time $T_{A\cup B}$ has no impact. However, in our numerical experiments we find that use of stopped trajectories improves the accuracy of \eqref{Rate_estimator} and, in particular, \eqref{eq:estimator_JAB} below, for most choices of $t$. Given an estimate of $R_{AB}$, the rate constant is
\begin{equation}\label{Rate_estimator2}
    k_{AB}=\frac{R_{AB}}{\sum_iq_-(X^{(i)}(0))w(X^{(i)}(0))}.
\end{equation}
The denominator in \ref{Rate_estimator2} is the mean of the backward committor, which is the fraction of time the system spends having last visited state A.

As noted above, we can also use simulations to understand how reactive trajectories flow through a CV space.  One way to do this is to partition the space into discrete states and then estimate the reactive fluxes between pairs of states\cite{metzner_transition_2009}. However, the resulting directed graph can be complicated and difficult to interpret.  When the sample paths are continuous the reactive flux between neighboring values in CV space is can be summarized as a single vector field in CV space. If $\theta$ is a vector-valued CV and $ds$ is a bin in CV space of volume $\lvert ds \rvert$, the reactive current at point $s$ is
\begin{equation}\label{eqn:def_JAB}
\begin{split}
    J_{AB}^{\theta}(s)= \lim_{t,\, \lvert ds \rvert\rightarrow 0}\frac{1}{2t\, \lvert ds \rvert} \int& 
    \left(\mathcal{T}^{t}[\theta q_+](x) - \theta(x)\mathcal{T}^{t} q_+(x)\right)
    \1_{\{\theta\in ds\}}(x) q_-(x) \pi(dx) 
    \\
    +&\left(\mathcal{T}^{t}[\1_{\{\theta\in ds\}}\theta q_+](x) - \theta(x)\mathcal{T}^{t} [\1_{\{\theta\in ds\}}q_+](x)\right)
   q_-(x) \pi(dx)
\end{split}
\end{equation}
 In appendix \ref{secn:cvcurrent}, we show that $J_{AB}^{\theta}(s)=\int J_{AB}\cdot \nabla \theta(x)\delta(\theta(x)-s)\pi(dx)$, and we establish that the projected reactive current satisfies
 \begin{equation}\label{eqn:def1}
\int_{\partial C^{\theta}}J^{\theta}_{AB}(s)\cdot n_{C^{\theta}} d\sigma_{C^{\theta}}=\int_{\partial C} J_{AB}\cdot n_C d\sigma_C,
\end{equation}
where $C^{\theta}$ is any region of CV space such that its inverse image (under the CV mapping) in the full configuration space, $C$, contains $A$ and does not intersect $B$.  To estimate $J_{AB}^{\theta}$ from trajectory data we have the following estimator:
\begin{multline}\label{eq:estimator_JAB}
    J^\theta_{AB}(s) \approx \frac{1}{2t\lvert ds \rvert}\sum_{i=1}^M  q_+(X^{(i)}(t\wedge T_{A\cup B})) \left(\theta(X^{(i)}(t\wedge T_{A\cup B}))-\theta(X^{(i)}(0))\right)\\
    \times \1_{\theta \in ds}(X^{(i)}(0))q_-(X^{(i)}(0))
     w(X^{(i)}(0))\\
     + \frac{1}{2t\lvert ds \rvert}\sum_{i=1}^M  q_+(X^{(i)}(t)) \left(\theta(X^{(i)}(t))-\theta(X^{(i)}(S_{A\cup B}(t)))\right)\\
    \times \1_{\theta \in ds}(X^{(i)}(t))q_-(X^{(i)}(S_{A\cup B}(t)))
     w(X^{(i)}(0))
\end{multline}
Note that the lag time $t$ in \eqref{Rate_estimator} and \eqref{eq:estimator_JAB} need not be the same as the lag time $\tau$ used to estimate the committors $q_+$ and $q_-$.  In contrast to the role of $\tau$, even with perfect sampling and a perfect basis, estimates of TPT quantities will depend on $t$.  Several considerations are involved in the choice of $t$.  For larger values of $t$ \eqref{Rate_estimator} and \eqref{eq:estimator_JAB} incur significant bias due to poor approximation of the $t\rightarrow 0$ limit in \eqref{eq:Nint} and \eqref{eqn:def_JAB}.  On the other hand, for larger values of $t$, we found that \eqref{eq:estimator_JAB} suffers large statistical errors.  The choice of $t$ is additionally constrained by the need for the change of measure at all initial time points to be used in  \eqref{eq:estimator_JAB}, which requires that one uses the same lag time for computing both $J_{AB}^\theta$ and $w$.  A full analysis of the error sources is beyond the scope of this work, and in practice we choose a lag time that gives reasonable results for the change of measure and reasonable smoothness in the vector field.

\section{Simulation methods and choices}

In this section, we specify the computational procedure to generate and analyze the dataset for the unfolding and folding of trp-cage. We describe preparing the system and its underlying dynamics (Section \ref{secn:setup}), choosing collective variables based on their ability to distinguish metastable states (Section \ref{secn:cvs}), generating and validating the dataset of short trajectories (\ref{secn:dataset}), and defining the unfolded and folded states (Section \ref{secn:states}).

\subsection{System setup}
\label{secn:setup}
Unless otherwise noted, all molecular dynamics simulations were performed with GROMACS 5.1.4 \cite{Abraham2015} and PLUMED 2.3 \cite{Bonomi2009, Tribello2014, PLUMED2019} using the CHARMM36m force field \cite{MacKerell1998,Best2012,Huang2016} in the NVT ensemble at 300 K using the Langevin thermostat with a temperature coupling constant of 10 ps$^{-1}$ applied to all atoms, and a time step of 2 fs.  Bonds to hydrogen atoms were constrained using the LINCS algorithm \cite{hess_lincs_1997}.  Electrostatic interactions were computed using particle-mesh Ewald summation with a cutoff of 1.2 nm.  Lennard-Jones interactions were switched off from 1.0 to 1.2 nm using the default GROMACS switching function.  

The system was prepared from an NMR structure of trp-cage (PDB code 1L2Y \cite{neidigh_designing_2002}).  The protein was solvated in a 50 \AA\ cubic box with the TIP3P water model \cite{Jorgensen1983} using CHARMM-GUI 3.0 \cite{Jo2008, Lee2016}.  10 K$^+$ and 11 Cl$^-$ ions were added, bringing the system to charge neutrality and 150 mM KCl.  The energy of the system was minimized until the maximum force was below 1000 kJ/mol nm.  The system was then equilibrated for 1 ns in the NVT ensemble with position restraints (using a 1 fs timestep), 10 ns in the NPT ensemble with harmonic restraints on non-hydrogen atom positions (force constant 400 kj/mol nm$^2$ for backbone atoms and 40 kj/mol nm$^2$ for side chain atoms.) and a Parrinello-Rahman barostat with a pressure coupling constant of 5 ps$^{-1}$, 5 ns in the NPT ensemble without position restraints, and then 10 ns in the NVT ensemble without position restraints.  The cubic box length was determined from the restraint-free NPT equilibration run to be 4.48 nm and fixed at that value after that run. 

\subsection{Choice of CVs}
\label{secn:cvs}
The performance of DGA rests on having a dataset with good sampling of all states that contribute to the reaction mechanism.  As mentioned in the Introduction, the available physically weighted molecular dynamics data for trp-cage \cite{lindorff-larsen_how_2011} contain few unfolding and folding transitions.  We thus sought to use enhanced sampling methods to generate a dataset with improved representation outside the stable states.  To this end, we evaluated CVs for their ability to control sampling and resolve the unfolded and folded states.
 
 Based on previous studies \cite{juraszek_rate_2008,sidky_high-resolution_2019}, we considered five CVs:
 \begin{enumerate}
 \item The radius of gyration of the C$_\alpha$ atoms ($R_g$); 
 
 \item The root mean squared deviation (RMSD) of all C$_\alpha$ atoms from their positions in an equilibrated structure (RMSD$_{\rm full}$);
 
 \item The RMSD of the C$_\alpha$ atoms of residues 2 to 9, which make up the $\alpha$ helix in the native state (RMSD$_{\rm hx}$);
 
  \item The RMSD of the C$_\alpha$ atoms of residues 11 to 15, which make up the 3-10 helix  in the native state (RMSD$_{\rm 3-10}$);
 
 \item The end-to-end distance ($d$).
 \end{enumerate}
 $R_g$, RMSD$_{\rm full}$, and RMSD$_{\rm hx}$ were used in ref.\ \citenum{{juraszek_rate_2008}}, and RMSD$_{\rm 3-10}$ was used in ref.\ \citenum{{sidky_high-resolution_2019}} (there defined only to residue 14), where they found that it was able to resolve several metastable states identified by spectral clustering.
  
To explore how these collective variables change as trp-cage unfolds, we ran a series of Adiabatic Bias Molecular Dynamics (ABMD) \cite{Marchi1999} simulations to drive unfolding from the equilibrated native structure. ABMD uses a ratchet-and-pawl-like bias to trap spontaneous fluctuations that move the system forward in selected CVs. By applying ABMD with different combinations of the CVs above, we found that RMSD$_{\rm full}$ and RMSD$_{\rm 3-10}$ yielded reasonable control of the system and enabled exploration of all metastable states characterized in previous studies. 

\subsection{Generation of the DGA dataset}
\label{secn:dataset}

\begin{figure}[h!]
\includegraphics[scale=.5]{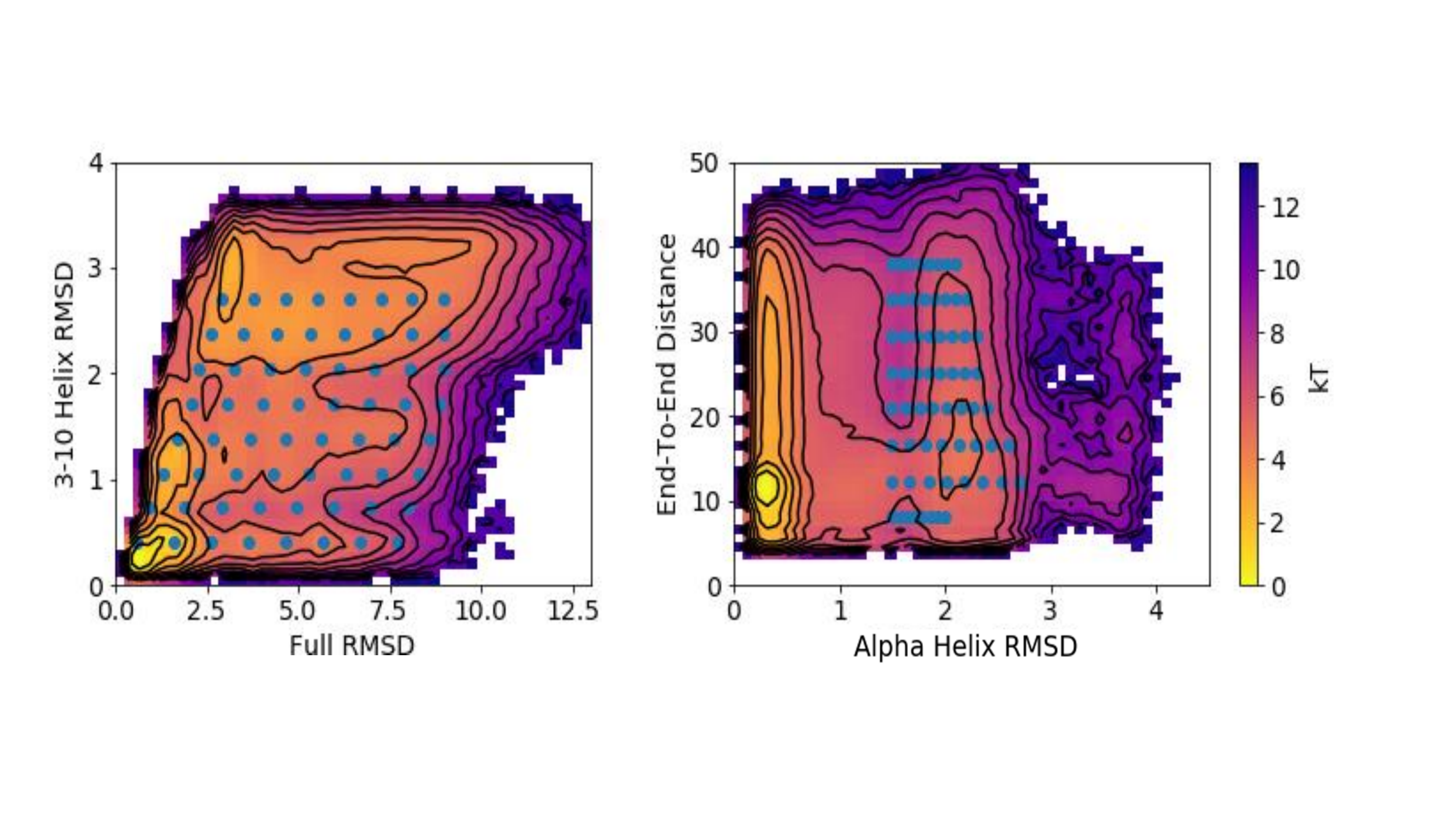}
\caption{\label{fig:abmd}
Initialization points for the dataset of short trajectories.  ABMD targets (symbols) are overlaid on DGA PMFs (color scale and contours, spaced every 1 $k_BT$) for the CVs used for steering. (left) The initial 64 ABMD targets were based on RMSD$_{\text{full}}$ and RMSD$_{\text{3-10}}$; 14 free simulations of length 30 ns were launched from each of the structures resulting from these ABMD simulations.  (right) 64 ABMD targets in RMSD$_{\text{hx}}$ and end-to-end distance added to ensure adequate sampling of the unfolded state; 2 free simulations of length 30 ns were launched from each of the structures resulting from these ABMD simulations.}
\end{figure}

To initialize a dataset of short trajectories for DGA, we defined a grid of 64 points in the space of RMSD$_{\rm full}$ and RMSD$_{\rm 3-10}$ (Figure \ref{fig:abmd}).  We then used 64 independent ABMD simulations to steer the system to each of these points from the final structure from the equilibration simulations described in Section \ref{secn:setup}.  We ran each ABMD simulation for 1 ns, saving the structure every 5 ps; the force constants were 1.25 kJ/(mol \AA$^2$) and 1.0 kJ/(mol \AA$^2$) for RMSD$_{\rm full}$ and RMSD$_{\rm 3-10}$, respectively.  From the set of all recorded structures, we chose the 64 structures closest to the targets and equilibrated each for 1 ns with a harmonic restraint with the same force constants as in the AMBD simulations. From each of the resulting structures, we then launched 14 free simulations (with different random number generator seeds) of length 30 ns each, saving structures every 5 ps.  

From this dataset, we computed all possible two-dimensional potentials of mean force (PMFs) involving the CVs listed in Section \ref{secn:cvs}. We compared these PMFs with corresponding ones from replica exchange umbrella sampling (REUS). Based on the DGA PMFs, we used the RMSD of the $\alpha$ helix (RMSD$_{\rm{hx}}$), and the RMSD of the 3-10 helix (RMSD$_{\rm{3-10}}$), and the end-to-end distance (\textit{d}) to control the sampling. REUS window centers were placed on a uniform $8\times 8\times 8$ grid of these three CVs, with RMSD$_{\rm{hx}}$ ranging from 0.3 to 2.8 \AA, RMSD$_{\rm{3-10}}$ ranging 0.3 to 3.3 \AA, and $d$ ranging from 6 to 38 \AA.  This grid fully covered the relevant areas of CV space identified by previous simulations. The force constants for the harmonic potentials for each window were 29.2 kJ/(mol $\cdot$ \AA$^2$) for RMSD$_{\rm{hx}}$,  20.3 kJ/(mol $\cdot$ \AA$^2$) for RMSD$_{\rm{3-10}}$, and 0.178 kJ/(mol $ \cdot $ \AA$^2$) for $d$, following ref.\ \citenum{Park2012}. To initialize each window, structures were taken from the DGA database that were closest to each window center. The built-in replica exchange functionality of GROMACS was used to create a three-dimensional replica exchange procedure, where structures from nearby windows were periodically exchanged \cite{Sugita2000}. Every window was first simulated for 100 ps, with swaps attempted between adjacent windows in $d$ space (i.e., window centers with the same RMSD$_{\rm{hx}}$ and RMSD$_{\rm{3-10}}$ values, but neighboring $d$ values) every 10 ps. This was repeated for a total of three 100 ps iterations, with the second and third iterations proposing swaps between neighboring windows in RMSD$_{\rm{hx}}$ and RMSD$_{\rm{3-10}}$, respectively. This 300-ps procedure was repeated until a total simulation time of 10 ns was reached for each window, with structures saved every 10 ps. Following this protocol, structures were exchanged across all of the three-dimensional grid, with exchange probabilities in the range 10-60\%. The PMF was constructed by using the Eigenvector Method for Umbrella Sampling (EMUS) \cite{Thiede2016} extended to REUS \cite{Antoszewski2020}. The REUS simulations were run until the asymptotic variance of the PMF dropped below 0.1 $(k_BT)^2$ (Figure \ref{fig:USAvars}).  

\begin{figure}[h!]
\begin{center}
\includegraphics[scale=.4]{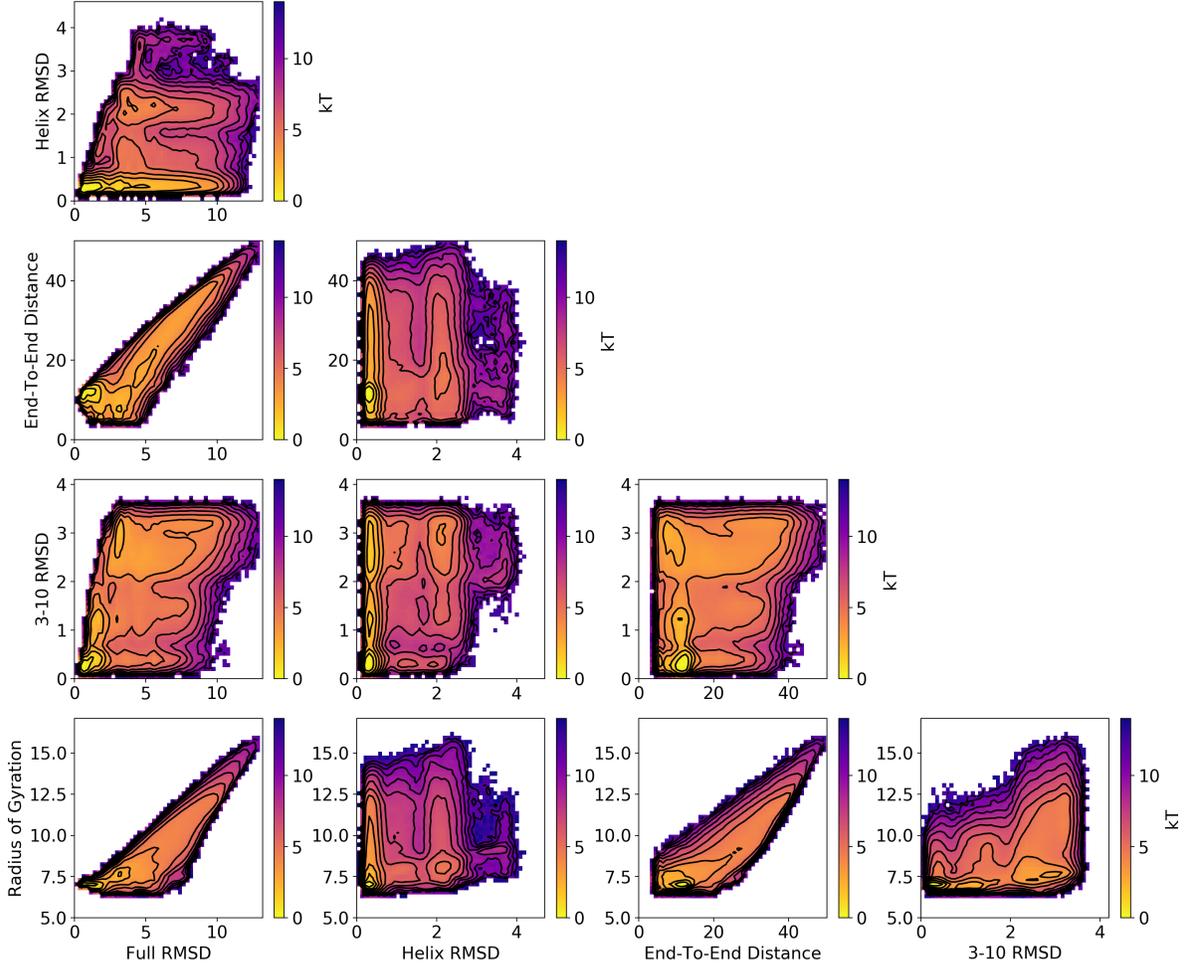}
\end{center}
\caption{\label{fig:triangle}
PMFs for the indicated CVs.  Results shown are computed by DGA with the modified distance basis set and a lag time of 0.5 ns. We use a $50\times 50$ grid to compute each PMF.  Similar results are obtained with other basis sets and REUS; see Figures \ref{fig:USminusInd}, \ref{fig:USminusTICA}, and \ref{fig:USminusSmooth}.}
\end{figure}

The REUS PMFs suggested that the initial DGA dataset did not adequately sample configurations with $\rm RMSD_{hx}>1.5\ \AA$ (Figure \ref{fig:USminusDGAInitialSampling}, note the lack of sampling toward the upper right areas of the plots compared with those in Figure \ref{fig:triangle}).  In this case several of the basins are missing, and the RMSD over all bins is $> 1.3\ k_BT$.  Therefore, we selected 64 more points from a grid with $\rm RMSD_{hx}>1.5\ \AA$ and a range of end-to-end distances from our short trajectory dataset. From each of these points, we released two new free molecular dynamics simulations of length 30 ns (Figure \ref{fig:abmd}B). With these additional trajectories, we obtained good agreement between DGA and REUS PMFs.  Adding the extra sampling improved the PMFs involving RMSD$_{\rm{hx}}$ the most, but other PMFs were also noticeably improved.  The dataset used for all further DGA calculations thus contains a total of 1024 trajectories, each of length 30 ns, with structures saved every 5 ps.

\subsection{State definitions}
\label{secn:states}
We found that PMFs projected onto only global measures of unfolding (RMSD$_{\rm full}$, $R_g$, and $d$) did not have clearly identifiable unfolded basins (Figures \ref{fig:abmd} and \ref{fig:triangle}).  By contrast, the PMF on the CVs tracking secondary structure (RMSD$_{\rm hx}$ and RMSD$_{\rm 3-10}$) had clearly identifiable unfolded and folded basins, as well as several intermediates.  Based on this analysis, we took the unfolded state to be
\begin{equation}
\frac{\rm |RMSD_{hx}-2.15\ \AA|^3}{\rm 0.008\ \AA^3}+\frac{\rm |RMSD_{310}-2.8\ \AA|^3}{\rm 0.125\ \AA^3}<1.  
\end{equation}
The folded state is
\begin{equation}
\frac{\rm (RMSD_{hx}-0.3\ \AA)^2}{\rm 0.0289\ \AA^2}+\frac{\rm (RMSD_{310}-0.3\ \AA)^2}{\rm 0.04\ \AA^2}<1\ {\rm and}\ d<17\ {\rm \AA}.
\end{equation}
We included the end-to-end distance constraint on the folded state to exclude structures which are extended but have the secondary structure intact. 

\begin{figure}[h!]
\includegraphics[scale=.5]{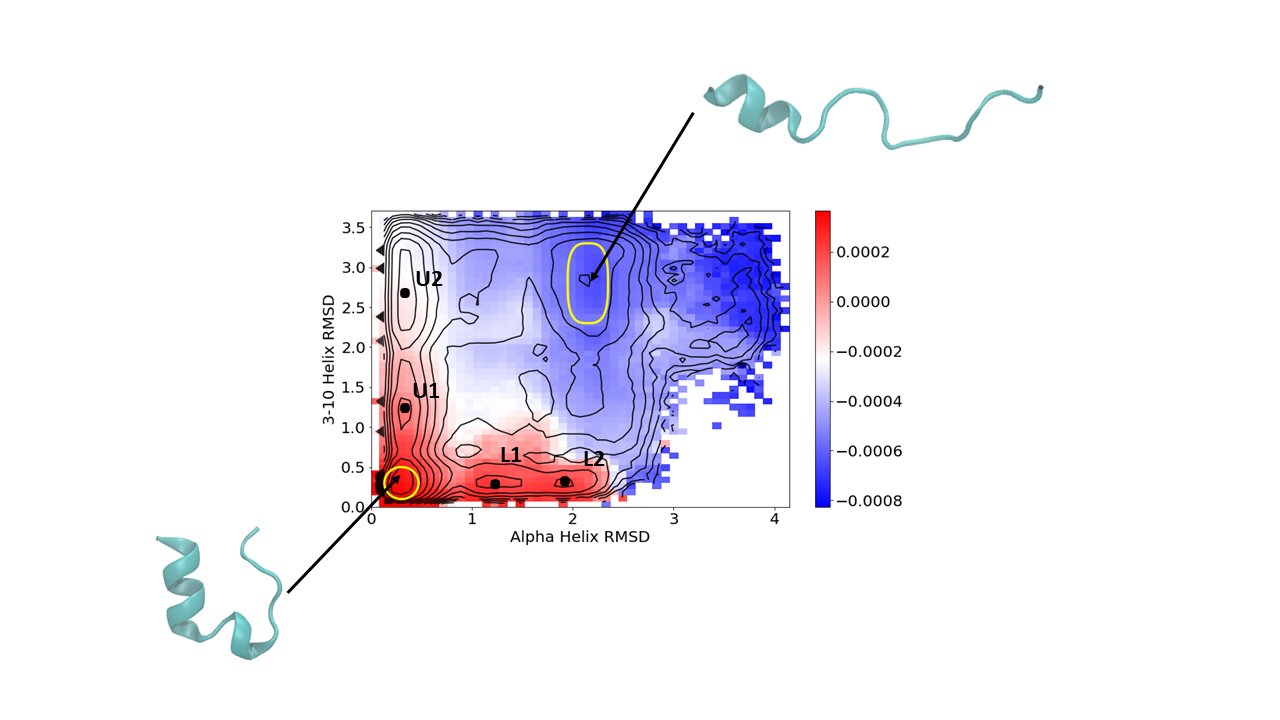}
\caption{\label{fig:TICA}
Top nontrivial TICA eigenvector averaged on the RMSD$_{\text{hx}}$ and RMSD$_{\text{3-10}}$ CVs with physical weighting.  The unfolded and folded states are indicated in yellow with representative structures.  Intermediate states in Table \ref{Tab::CVs} are marked and labeled.}
\end{figure}

Heterogeneous structures contribute to the unfolded state, making it challenging to define, and there is no guarantee that the choices above are optimal in any sense.
Because we expect unfolding and folding to be among the slowest motions of the system, an alternative would be to define the states in terms of the slowest mode of the system identified by a dimensionality-reduction algorithm. However, data-driven state definitions are often difficult to interpret physically, despite their theoretical justifications.  Furthermore,  data-driven state definitions can be difficult to incorporate into sampling algorithms.  We thus use physical CVs for path sampling, stratification, and state definitions, and we then check for consistency with a data-driven state choice.  

Figure \ref{fig:TICA} shows that the slowest mode of the system identified by TICA applied to the DGA dataset correlates with the PMF and switches between low and high values in going between the unfolded and folded states.  Here and going forward, all functions we project onto CVs are conditional averages of the form $\int f(x)\delta(\theta(x)-s)\pi(dx) / \int \delta(\theta(x)-s)\pi(dx)$.  We estimate these by binning our CV space into bins, and for each bin $ds$, plotting:
\begin{equation}\label{eqn:fxnplot}
   \frac{\int f(x)\delta(\theta(x)-s)\pi(dx)}{ \int \delta(\theta(x)-s)\pi(dx)}\approx \frac{\sum_i f(X^{(i)}(0)w(X^{(i)}(0))\1_{\theta\in ds}((X^{(i)}(0))}{\sum_i w(X^{(i)}(0))\1_{\theta\in ds}((X^{(i)}(0))}.
\end{equation}
We furthermore show in Section \ref{secn:committors} that this mode correlates with the committor.  We thus feel that RMSD$_{\rm hx}$ and RMSD$_{\rm 3-10}$ enable the clearest two-dimensional projection of the reaction and present most of our results in terms of these CVs.  In addition to the unfolded and folded states, we define four intermediate states U1, U2, L1, and L2 shown on Figure \ref{fig:TICA}.  In the next section, we apply our DGA and TPT formalism to show that trp-cage can fold along an upper path through intermediates U1 and U2, or a lower path through L1 and L2.

\section{Trp-cage analysis}
\label{secn:trpcage}
In this section, we evaluate how the three basis sets described in Section \ref{secn:basis} (indicator functions of pairwise distances, indicator functions of TICA coordinates, and pairwise distances modified to satisfy the boundary conditions) impact the performance of DGA for estimating PMFs, rates, committors, and reactive currents for the unfolding and folding of the trp-cage miniprotein.  Where possible, we compare our results with references obtained by independent means.  

\subsection{Comparison of PMFs}
\label{secn:PMFs} 

Figure \ref{fig:triangle} shows PMFs computed on each pair of the physically motivated CVs with DGA with the modified distance basis set.  The corresponding PMFs from REUS are shown in Figure \ref{fig:REUS}; difference maps comparing the results obtained with the two methods and three basis sets are shown in Figures \ref{fig:USminusInd}, \ref{fig:USminusTICA}, and \ref{fig:USminusSmooth}.  All of the main basins identified by REUS are present in the DGA PMFs, and there is good quantitative agreement between REUS and DGA, with RMSDs of $<1\ k_BT$ for all three basis sets (that said, of these, the distance indicator basis set results in the largest deviations).  Consistent with their agreement with the REUS PMF, the three DGA PMFs are in agreement with each other.  We did observe that REUS tends to give slightly flatter PMFs than DGA with all three basis sets.  In principle, there are two sources of error in the DGA PMFs: (i) approximation error from representing the true change of measure with a basis expansion and (ii) estimation (sampling) error.  Analysis of error in DGA will be the subject of future work.   Error in US is discussed in refs.\ \citenum{Thiede2016}, \citenum{Antoszewski2020}, and  \citenum{dinner_stratification_2020}.

\begin{table}[ht]
\centering
\caption{\label{Tab::CVs}
CV values for metastable states.}
\begin{tabular}{|l|l|l|r|l|l|}
\hline
  State & RMSD$_\text{full} / \text{\AA}$ & RMSD$_\text{hx}/ \text{\AA}$ & $d / \text{\AA}$    & RMSD$_\text{3-10}/ \text{\AA}$ & $R_g$/ \text{\AA} \\ \hline\hline
Folded  & 1.1  & 0.30     & 11.1 & 0.30      & 7.0  \\ \hline
Unfolded  & 5.8  & 2.1      & 20.2 & 2.8       & 9.2  \\ \hline
U1 & 2.4  & 0.34     & 13.1 & 1.2       & 7.3  \\ \hline
U2 & 5.2 & 0.34     & 19.3 & 2.8       & 8.8  \\ \hline
L1 & 2.2  & 1.2      & 9.5  & 0.30      & 7.2  \\ \hline
L2 & 2.6  & 1.9      & 14.5 & 0.30      & 7.3  \\ \hline
\end{tabular}
\end{table}

\begin{figure}[h!]
\includegraphics[scale=.5]{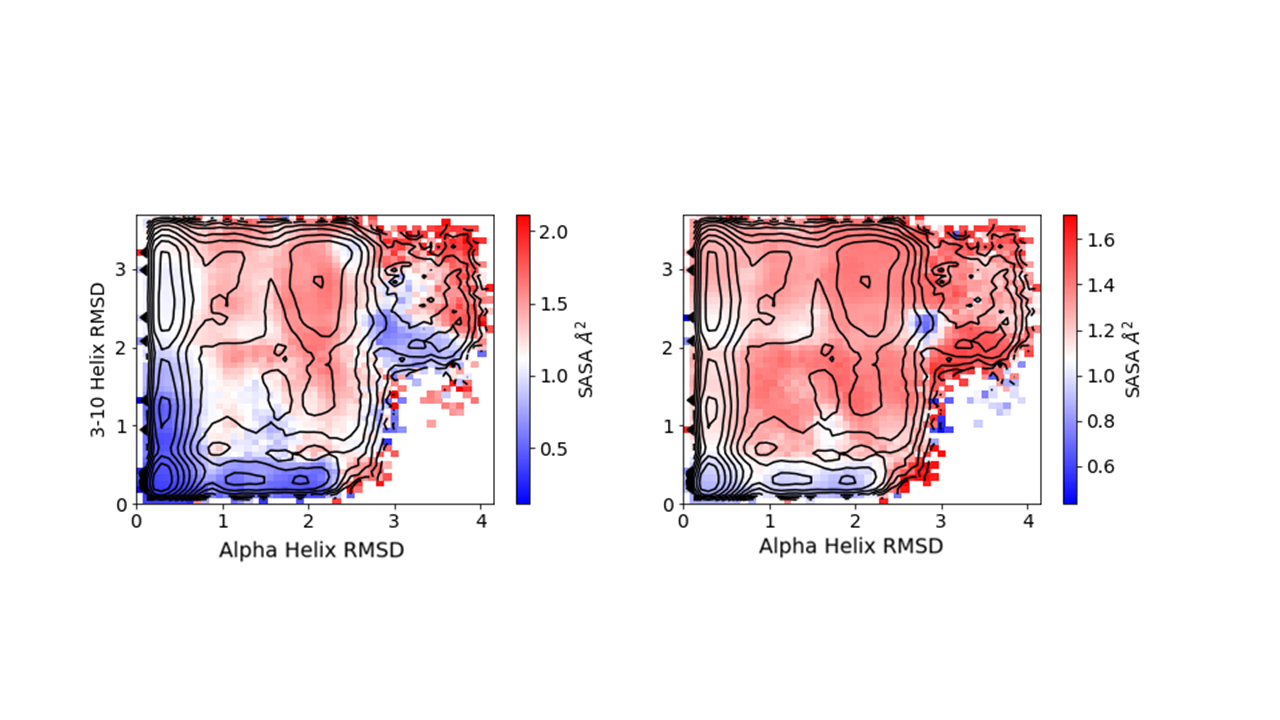}
\caption{\label{fig:sasa}
Equilibrium average solvent accessible surface area (SASA) projected onto the RMSD$_{\text{hx}}$ and RMSD$_{\text{3-10}}$ CVs for (left) trp-6 and (right) proline-12.}
\end{figure}

We found that the projection onto the RMSD$_{\text{hx}}$ and RMSD$_{\text{3-10}}$ coordinates was best able to separate the pathways and states of interest, so we now focus on this projection.  Figure \ref{fig:TICA} indicates the folded (lower left) and unfolded (upper right) basins, as well four intermediates.  The intermediates define two pathways, which we label upper (with intermediates U1 and U2) and lower (with intermediates L1 and L2).  Table \ref{Tab::CVs} gives the five CV values for each of the six states.  

To understand the characteristics of the intermediate states, we turn to Figure \ref{fig:sasa}, which shows the solvent-accessible surface area (SASA) of trp-6 on the left, and pro-12 on the right.  We find that the U2 intermediate state is characterized by partial solvation of the hydrophobic core, measured by the SASA of trp-6, and nearly full detachment of pro-12.  Furthermore, the U2 state is significantly more extended than the lower pathway intermediates as measured both by $R_g$ and end-to-end distance.  In addition to being more compact, with near-native $R_g$ values, L1 and L2 have near-native trp-6 and pro-12 SASA values, suggesting the hydrophobic core is fully formed.  These intermediate states can be mapped to those previously reported in the literature.  Bolhuis and Jurazek \cite{juraszek_sampling_2006} identified three folding intermediates.  Our U1 and U2 intermediates roughly map onto their P$_d$ and I intermediates, and our L1 and L2 intermediates roughly map onto their L intermediate.  U1 and U2 also correspond to states S$_7$ and S$_0$ identified by Sidky {\it et al.} \cite{sidky_high-resolution_2019}.

\subsection{Comparison of committors}
\label{secn:committors}

\begin{figure}[hbt!]
\includegraphics[scale=.9]{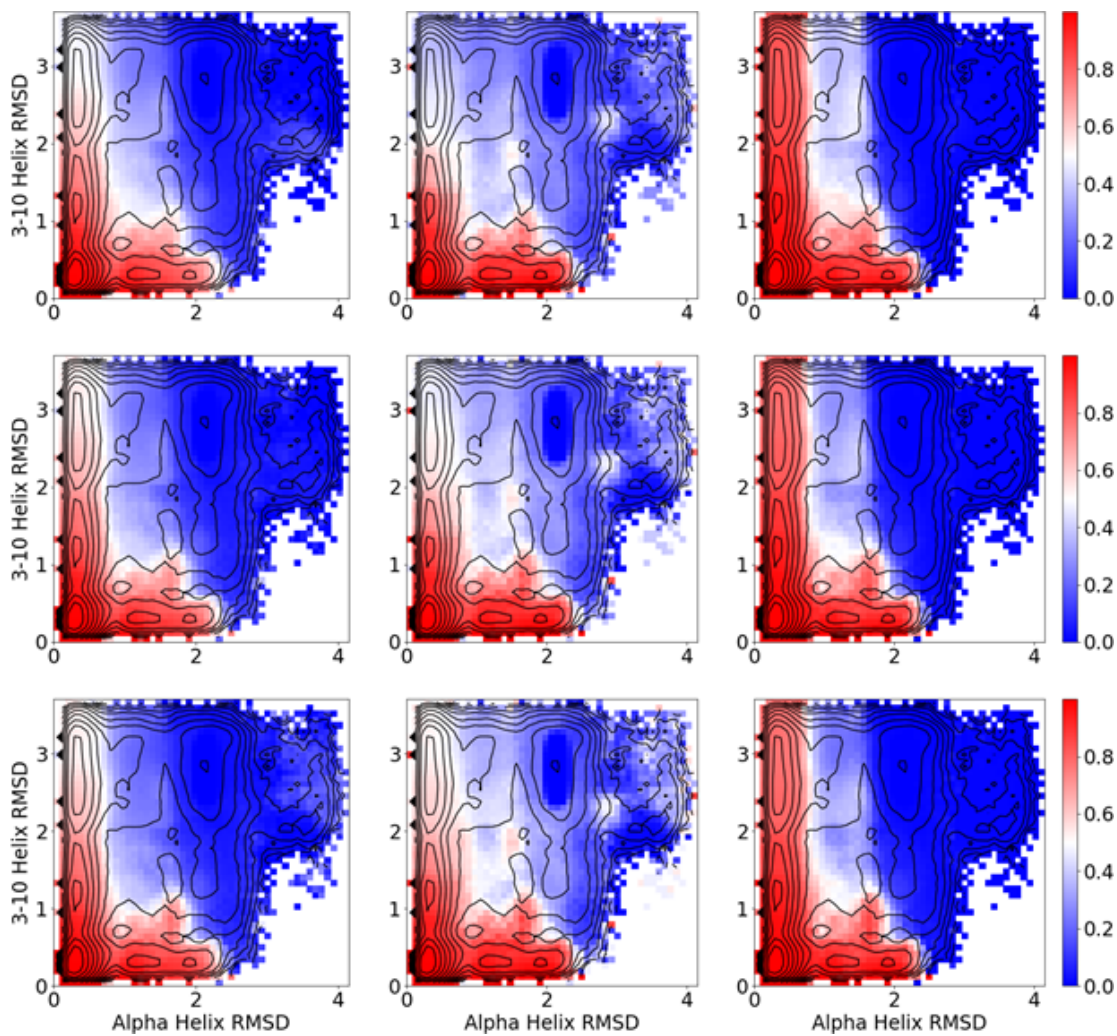}
\caption{\label{fig:Comms}
DGA forward committors. Left, middle, and right columns are computed with the modified distance, distance indicator, and TICA indicator basis sets, respectively.  Top, middle, and bottom rows are computed with lag times of 0.5, 2.5, and 7.5 ns, respectively.}
\end{figure}

We next calculated both forward and backward committors using DGA with the three basis sets and lag times ranging from 0.5 ns to 12 ns  (Figure \ref{fig:Comms} and Figure \ref{fig:backcomm}).   As they should, the backward committors mirror the forward committors, so we focus our discussion on the latter.  The timescale of trp-cage folding is on the order of 5 $\mu$s from both experiment \cite{qiu_smaller_2002} and simulation\cite{juraszek_rate_2008}, thus both our trajectory lengths (30 ns) and lag times are several orders of magnitude shorter than the motions of interest, providing an appropriate setting in which we expect DGA to show benefits. 

In contrast to the PMFs, we found the committors to be sensitive to the choice of basis set (and associated guess function).  The modified distance basis set, in addition to being substantially faster to construct as it avoids slow and unstable high-dimensional clustering, is less prone to discontinuities at the boundary than the distance indicator function basis set.  The TICA indicator function basis set performs similarly to the modified distance basis set and has the advantage over the distance indicator basis set that clustering on the lower-dimensional subspace is significantly faster and more stable.

For a given basis set, we found relatively little variation in the committors across lag times.  This is in contrast to variational approach for conformational dynamics (VAC) algorithm, where the results can strongly depend on the lag time \cite{webber_error_2020} (although this can be mitigated by using multiple lag times\cite{lorpaiboon_integrated_2020}).  We postpone a full investigation of DGA's error properties, and in particular its dependence on the choice of lag time, to future work.



Because we expect unfolding and folding to be among the slowest motions of the system, we can validate the DGA committors by comparing them with the slowest mode of the system identified by TICA.  Comparing Figures \ref{fig:TICA} and \ref{fig:Comms} shows that the largest TICA eigenvector (estimated with a lag time of 0.5 ns) correlates almost perfectly with the estimated committors obtained with the modified distance basis set, when projected onto RMSD$_{\rm hx}$ and RMSD$_{3-10}$.  
The agreement between these two independent calculations furthermore suggests that the physically motivated CVs capture the behavior detected by the data-driven method.  
In this projection, we see that the transition states fall where the SASA of trp-6 (Figure \ref{fig:sasa}) changes rapidly.

\begin{figure}[h!]
    \centering
    \includegraphics[scale=.5]{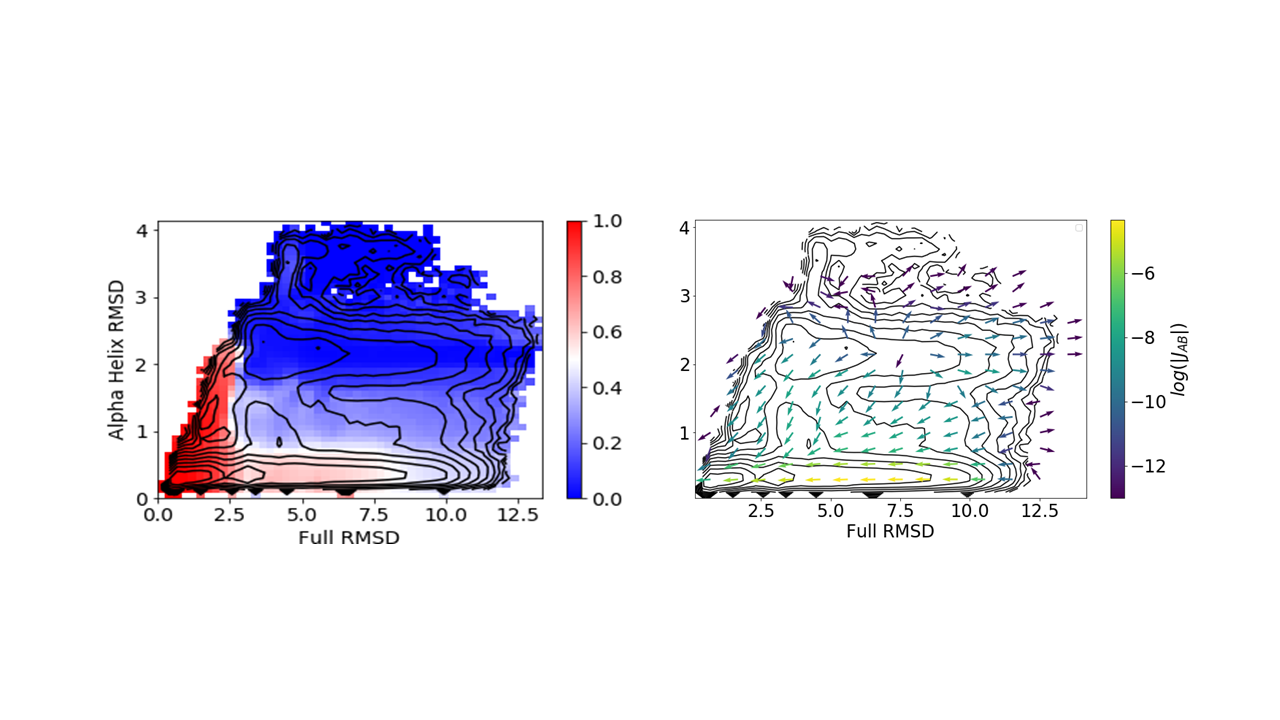}
    \caption{\label{fig:Bolhuis}
    Forward committor (left) and  reactive current (right) projected onto the  RMSD$_{\rm hx}$ and full RMSD CVs used in ref.\ \citenum{juraszek_sampling_2006}.  Results shown are computed with  the modified distance basis set and a lag time of 0.5 ns.}
    \label{fig:my_label}
\end{figure}

As an additional validation, we used DGA with the modified distance basis set and a lag time of 0.5 ns  (Figure \ref{fig:Bolhuis}) to compute committors on the CVs used by Juraszek and Bolhuis \cite{juraszek_sampling_2006}. 
When projected onto RMSD and RMSD$_{\rm hx}$, the positions of the transition states in Figure 4 of ref.\ \citenum{{juraszek_sampling_2006}} fall in areas estimated to have $q_+=0.5$ (white in Figure \ref{fig:Bolhuis}).  The traditional shooting approach employed in ref.\ \citenum{juraszek_sampling_2006} is quite computationally costly and provides information about only a limited number of structures.  Our ability to capture the transition states thus makes clear the benefit of DGA. We discuss DGA's ability to provide mechanistic information further in the next section.

\subsection{Reactive currents}
\begin{figure}[h!]
\includegraphics[scale=.5]{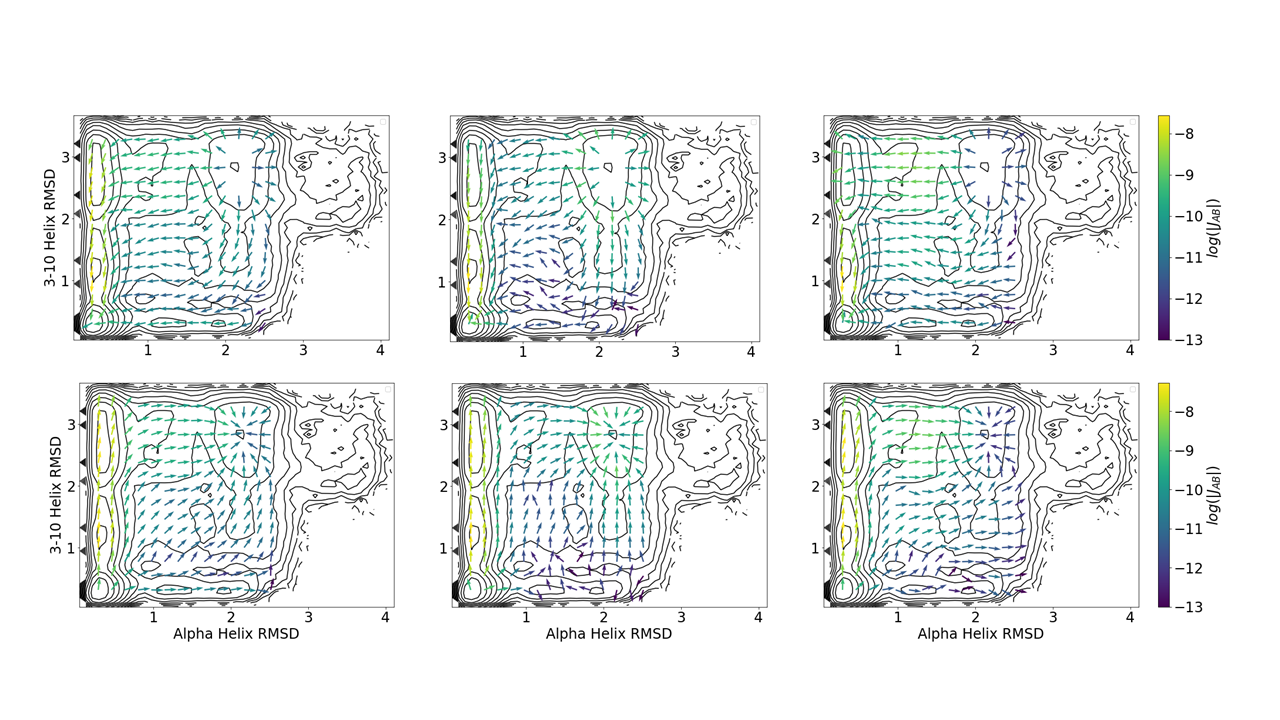}
\caption{\label{fig:Currents}
Folding (top) and unfolding (bottom) reactive current projected onto the RMSD$_{\text{hx}}$ and RMSD$_{\text{3-10}}$ CVs using \eqref{eq:estimator_JAB} with the three choices of basis set.  Left, middle, and right columns are computed with the modified distance, distance indicator, and TICA indicator basis sets, respectively.  All computations use a lag time of 0.5 ns.}
\end{figure}

We computed reactive currents for the three basis sets using the estimator in \eqref{eq:estimator_JAB} and the committors from the the previous section (Figure \ref{fig:Currents}).  For this calculation, we use the shortest lag time of 0.5 ns for both the committor and reactive current, though in principle they could be chosen separately.  As previously, we primarily present our results projected onto RMSD$_{\rm{hx}}$ and RMSD$_{\rm{3-10}}$.  Overall the results for the three basis sets are similar, though the distance indicator basis set exhibits greater noise around (RMSD$_{\rm{hx}}$, RMSD$_{\rm{3-10}}$) = (1.3 \AA, 1.3 \AA), consistent with the plateau in the committor in this region.

The currents, which provide information directly about dynamics, confirm the presence of two paths for the folding process: an upper path with formation of the $\alpha$ helix prior to formation of the 3-10 helix, and a lower path with the order of these events transposed.  The upper path proceeds through intermediates U1 and U2, with folding beginning with formation of the $\alpha$ helix and partial desolvation of trp-6, followed by full formation of the 3-10 helix.  The lower path proceeds through L1 and L2, with folding beginning with collapse into the L2 intermediate with no $\alpha$ helix, but the hydrophobic core fully formed, followed by formation of the $\alpha$ helix.  Both of these paths correspond to troughs in the PMFs on these CVs.  

Previous studies have found multiple pathways resembling the ones we find here.  Kim et al.\ \cite{kim_systematic_2015} used diffusion maps to identify two pathways: one with tertiary contacts forming first, followed by $\alpha$ helix formation, and another with the order transposed.  Jurazek and Bolhius came to similar conclusions using transition path sampling \cite{juraszek_sampling_2006}.

An advantage of the reactive current is that we can use it to assign weights to the two paths.  By computing the relative flux crossing RMSD$_{\text{3-10}}=1.8$ \AA\  with either RMSD$_{\text{hx}}<1.4$ \AA\ (upper pathway) or RMSD$_{\text{hx}}>1.4$ \AA\ (lower pathway), we conclude that 88\% of the reactive paths proceed by first forming the $\alpha$ helix, and then the 3-10 helix and hydrophobic core (i.e., the upper pathway).  Although we are not aware of a previous estimate of the reactive current for this system, we can compare these numbers to the frequencies with which transition path sampling sampled the pathways in ref.\ \citenum{{juraszek_sampling_2006}}.  There, Juraszek and Bolhuis observed the pathway in which tertiary contacts form first (i.e., the lower pathway) 80\% of the time.  The difference may be due to different CV and state definitions (Jurazek and Bolhuis \cite{juraszek_sampling_2006} used 5 CVs in their state definitions , whereas we consider only RMSD$_{\rm 3-10}$ and  RMSD$_{\rm hx}$) or force field and setup differences.  

\subsection{Rates}
Finally, we computed rates using the estimator in \eqref{eqn:bareRate_S}.  We present our results as inverse rates (unfolding and folding times) to make comparisons to lag times and trajectory lengths clear.  As mentioned previously, these times are expected to be on the order of microseconds.  In particular, Juraszek and Bolhuis used transition interface sampling to estimate inverse unfolding and folding rates of 1.2 $\mu$s and 0.4 $\mu$s \cite{juraszek_rate_2008}, though as noted previously those results are for a different model.  

\begin{figure}[h!]
\includegraphics[scale=.5]{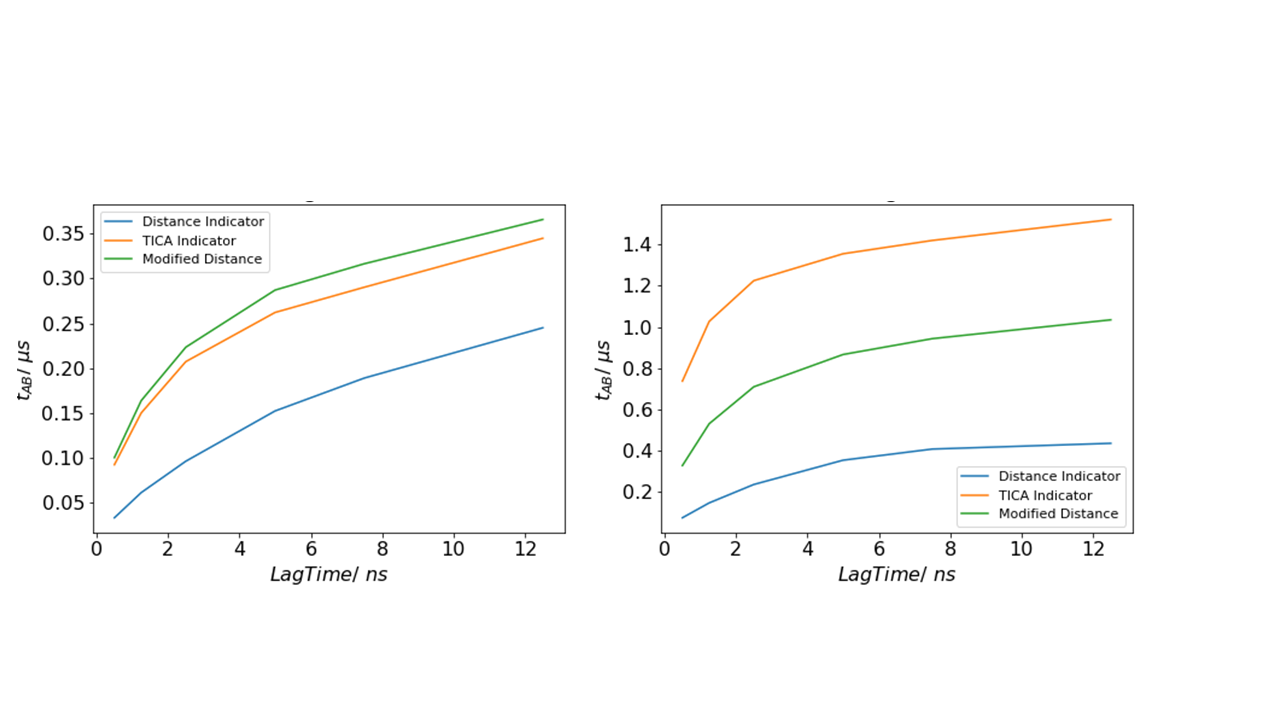}
\caption{\label{fig:Rates}
Inverse rates estimated for folding (left) and unfolding (right).}
\end{figure}

All three basis sets gave rate estimates that were within an order of magnitude of those numbers.  However, the results for the distance indicator basis were markedly faster.  Furthermore, in all three cases, the inverse rate exhibited significant dependence on lag time.  Our analysis of the trajectory of the K8A mutant suggests the need for a lag time of at least 100 ns (consistent with ref.\ \citenum{sidky_high-resolution_2019}), though as discussed in the Introduction, these data do not contain a sufficient number of unfolding and folding events to obtain accurate rate estimates.  Juxtaposed with the lack of sensitivity to lag time for the committor and reactive current, these observations suggest that DGA's strength is in its ability to give statistical insight into mechanisms with relatively little data, but that rates may be more efficiently computed by methods that directly sample relevant statistics such as stratification schemes\cite{dinner_trajectory_2018}.

\subsection{Demonstration of delay embedding}
\label{secn:delaydemo}

\begin{figure}[h!]
\includegraphics[scale=.5]{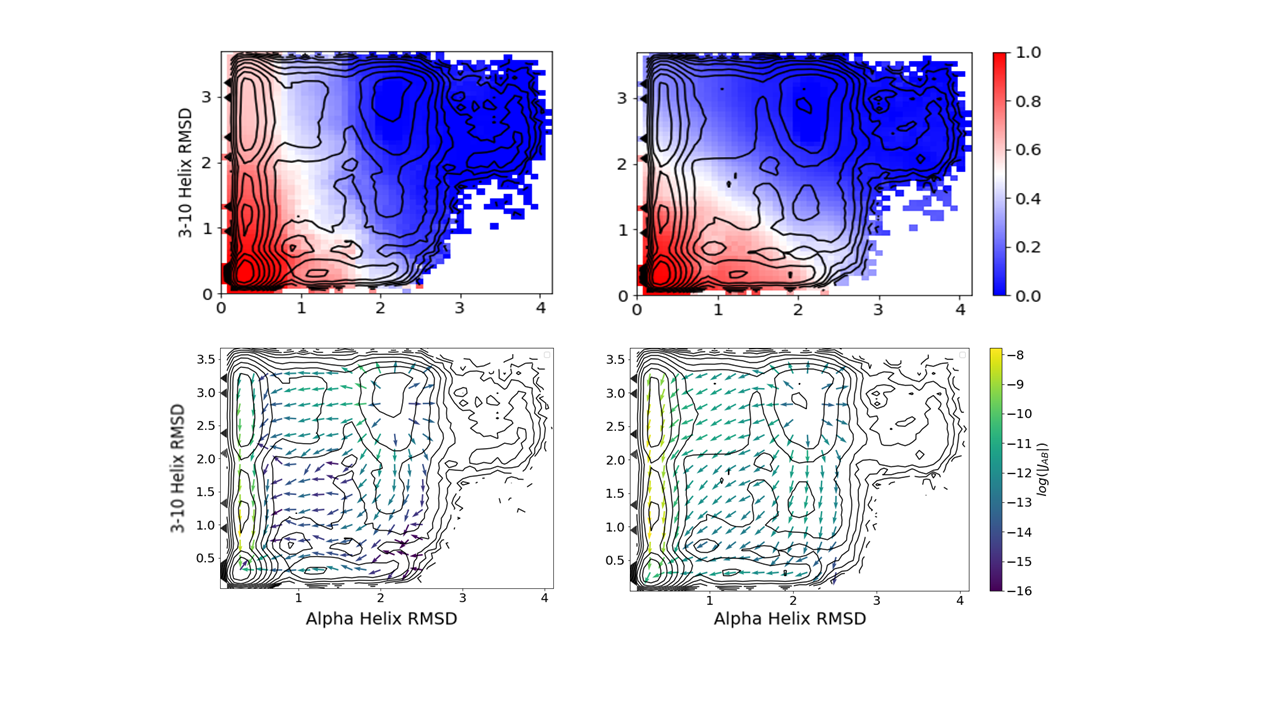}
\caption{\label{fig:DEB}
Comparison of DGA estimates for the forward committor (top) and reactive current for folding (bottom) with the modified distance basis set on a feature space restricted to the five physical CVs (right) and a delay-embedded feature space (left).  The delay-embedded results are obtained with a delay of $\delta=0.125$ ns, $N=40$ images, and a DGA lag time of 0.5 ns.}
\end{figure}

As described in Section \ref{secn:delayembedding}, delay embedding can be used to construct an approximately Markovian process when the feature space does not fully capture the dynamics.
To illustrate this idea using our trp-cage dataset, we restrict the feature space to the five physical CVs and apply DGA with the modified distance basis set on either the feature space itself or the delay-embedded feature space.  Figure \ref{fig:DEB} shows the reactive currents and committors resulting from DGA on these two spaces.  We find that the committor and current constructed from the delay embedded representation largely agree with the DGA result constructed on the 153 pairwise distances.  Without delay embedding, we find several qualitative disagreements, in particular the U2 state has a committor value close to zero, and the reactive current does not resolve the two pathways since many of the arrows point directly towards the folded state.

\FloatBarrier 

\section{Conclusions}
In this paper, we have cast the dynamical Galerkin approximation (DGA) \cite{thiede_galerkin_2019} for computing chemical kinetic statistics from short trajectories in terms of the stopped transition operator.  This formulation can be immediately translated into expressions that can be applied to simulation data.  It also clarifies the role of the lag time, showing that estimates of conditional expectations computed by DGA are exact in the infinite basis and data limit, independent of the choice of lag time.  

To evaluate DGA's performance, we generated and carefully validated a dataset of short trajectories for the unfolding and folding of the trp-cage miniprotein, a well-characterized system.  We used umbrella sampling to validate our short trajectory dataset by comparing the resulting PMFs.  Quantitative agreement between the PMFs was observed, suggesting that our short trajectory dataset had sufficient sampling to compute dynamical statistics.  The PMF calculations furthermore enabled us to rapidly assess different combinations of CVs for their abilities to separate metastable states.  The $\alpha$ helix RMSD and 3-10 helix RMSD in particular allowed us to resolve intermediates to a greater degree than found in previous studies.

We next applied DGA to compute forward and backward committors between the unfolded and folded states.  
We evaluated a number of competing estimators for the backward committor and found that one based on forward trajectories weighted by the stationary distribution gave the best results. 
The committors by themselves are not able to identify reaction pathways or transition states, but they can be combined according to transition path theory to extract this information.  Specifically, we introduce a new estimator for the TPT rate, and a projection formula and corresponding estimator for the reactive current in a CV space.  Our projected reactive current allows us to easily resolve and visualize the pathways that the system takes in arbitrary CV spaces, and even lets us assign relative weights to these pathways. Acquiring this kind of mechanistic information has previously been possible only through transition path sampling and related methods;  such methods do not as readily allow exploration of CVs and state definitions because the sampling is linked directly to them.

We introduced a simple procedure that takes an arbitrary set of molecular features and adapts them to produce a basis set that satisfies the homogeneous boundary conditions.  Using pairwise distances as the molecular features, we compared the performance of such a basis set with indicator functions on the molecular features and indicator functions on TICA coordinates.  Other basis constructions such as diffusion maps and radial basis functions are possible, and we expect that the best choice will be system dependent.  We applied our DGA and TPT formalism to our dataset, and identified intermediate states and pathways which have been previously reported in the literature, providing further validation of our methods.  We found that the estimates of the TPT rate, while on the same order of magnitude as previous estimates, nevertheless show significant dependence on lag time.  Finally, we showed that delay embedding can be an effective strategy for constructing a molecular representation with approximately Markovian dynamics from a low-dimensional feature space.

Our results suggest several interesting directions for future investigation.  We have seen that in our trp-cage application the choice of lag time has only a modest effect on DGA estimates of conditional expectations, while TPT quantities, and in particular the rate depend sensitively on lag time. Recently, we showed that integrating over lag times for VAC improves the robustness of that method \cite{lorpaiboon_integrated_2020}.  It will be interesting to see if an analogous strategy can improve rate estimates from DGA. An in depth mathematical study of DGA's error and its dependence on lag time along the lines of our previous analysis of VAC \cite{webber_error_2020} is also in order.  Though DGA has performed well in our tests so far, looking ahead to larger and more complex systems, it may become necessary  move away from a Galerkin approach and toward more flexible representations of the kinetic functions we seek to approximate.  This would be consistent with the current trend toward using neural networks to represent eigenfunctions in spectral estimation \cite{mardt_vampnets_2018,sidky_high-resolution_2019}.  Introducing this higher level of representational flexibility while maintaining the reliability we observe in our trp-cage application of DGA will be a challenge.

\begin{acknowledgement}

We thank Erik Thiede and Justin Finkel for their critical readings of the manuscript and helpful feedback as well as D.\ E.\ Shaw Research for making available the K8A mutant trajectory.  We also than Robert Webber for helpful conversations.
Research reported in this publication was supported by the National Institute of General Medical Sciences of the National Institutes of Health under award number R35GM136381. Simulations were performed on resources from the Research Computing Center at the University of Chicago.

\end{acknowledgement}

\begin{suppinfo}
Mathematical appendices, asymptotic variances for REUS, and difference maps between DGA and REUS PMFs.
\end{suppinfo}

\bibliographystyle{unsrt}
\bibliography{references,trp_cage_bib,dga_paper}
\makeatletter\@input{suppaux.tex}\makeatother
\end{document}


\beginsupplement
\appendix
\section{Backward-in-time inner products}
\label{secn:innerproductderivation}
In this appendix we provide an elementary derivation of \eqref{eq:backip} which is key to our estimates of inner products involving the stopped backward-in-time transition operator $\mathcal{T}^t_{A\cup B}$. For the purposes of this derivation we assume that $X(t)$ is a discrete time process (so that $t$ is a non-negative integer) with probability density $p(x(1)|x(0))$ for transition from $x(0)$ to $x(1)$ and stationary density $\pi$.  The steady state backward-in-time process $X(-t)$ then has transition density
\begin{equation}
q(x(0)|x(1)) = \frac{p(x(1)|x(0))\pi(x(0))}{\pi(x(1))}.
\end{equation}
From this expression we immediately find that
\begin{multline}
\pi(x(t))q(x(t-1)|x(t))q(x(t-2)|x(t-1))\cdots q(x(0)|x(1))\\
= \pi(x(0))p(x(t)|x(t-1))p(x(t-1)|x(t-2))\cdots p(x(1)|x(0))\quad\quad
\end{multline}
relating the steady state backward-in-time path density to the steady state forward-in-time path density.  Therefore, for any path function $F(x(0),x(1),\dots,x(t))$
and any density $\mu$ (equivalent to $\pi$) we find (recalling that here $w = \pi/\mu$) that
\begin{align}
&\int \mathbb{E}\left[F(X(0),X(-1),\dots,X(-t))\,|\,X(0)=x\right] \mu(x)dx \notag \\
&\hspace{1.5cm}= \int  \frac{F(x(0),\dots,x(-t))}{w(x(0))} \pi(x(0))q(x(-1)|x(0))\cdots q(x(-t)|x(-t+1))dx(0)\cdots dx(-t) \notag \\
&\hspace{1.5cm}= \int  \frac{F(x(t),\dots,x(0))}{w(x(t))} \pi(x(t))q(x(t-1)|x(t))\cdots q(x(0)|x(1))dx(t)\cdots dx(0) \notag \\
&\hspace{1.5cm}= \int \mathbb{E}\left[\frac{F(X(t),X(t-1),\dots,X(0))}{w(X(t))}\,\middle|\,X(0)=x\right]w(x) \mu(x)dx\label{eq:fbpath}.
\end{align}

We will use \eqref{eq:fbpath} to find an expression for
\begin{equation}
\langle g, \mathcal{T}^{-t}_{A\cup B}f\rangle = \int  \mathbb{E}\left[g(x)
f(X(-T^-_{A\cup B}\wedge t))\,|\,X(0)=x\right]\mu(x) dx
\end{equation}
in terms of the forward-in-time process.  If
we choose
\begin{equation}
F(X(0),X(-1),\dots,X(-t)) = g(X(0))f(X(-T^-_{A\cup B}\wedge t)),
\end{equation}
then
\begin{equation}
\langle g, \mathcal{T}^{-t}_{A\cup B}f\rangle = \int  \mathbb{E}\left[F(X(0),X(-1),\dots,X(-t))\,|\,X(0)=x\right]\mu(x) dx.
\end{equation}
In terms of the forward process 
\begin{equation}
F(X(t),X(t-1),\dots,X(0)) = f(X(S_{A\cup B}(t)))g(X(t)),
\end{equation}
where we remind the reader of \eqref{eqn:saub}:
\begin{displaymath}
S_{A\cup B}(t) = \sup\{s\leq t: \, X(s)\in A\cup B\}.
\end{displaymath}

Applying \eqref{eq:fbpath} with this choice of $F$ yields \eqref{eq:backip}:
\begin{equation*}
\left\langle g, \mathcal{T}^{-t}_{A\cup B}\,f\right\rangle =
\int \mathbb{E}\left[f(X(S_{A\cup B}(t)))\frac{g(X(t))}{w(X(t))}
\,\bigg{|}\,X(0)=x\right]w(x) \mu(dx).
\end{equation*}


\section{A formula for the reactive current}
It has been shown\cite{vanden2006transition} that for a diffusion
with generator
\begin{equation}
\mathcal{L}f(x) = \sum_j b_j(x) \frac{\partial f}{\partial x_j}(x)
+ \frac{1}{2}\sum_{ij} a_{ij}(x) \frac{\partial^2 f}{\partial x_i \partial x_j}(x)
\end{equation}
the reactive current is the vector field given by
\begin{eqnarray}\label{eqn:JAB}
(J_{AB})_i &= &q_+(x)q_-(x)J_i+\nonumber\\
&&\pi(x)q_-(x)\sum_ja_{ij}(x)\frac{\partial q_+}{\partial x_j}(x)-\pi(x)q_+(x)\sum_ja_{ij}(x)\frac{\partial q_-}{\partial x_j}(x),
\end{eqnarray}
where $J$ is the equilibrium current:
\begin{equation}
J_i = \pi(x)b_i(x)-\sum_j\frac{\partial[\pi a_{ij}]}{\partial x_j}(x).
\end{equation}
To project the current onto a CV space of interest, we take the dot product with $\nabla \theta$ for any smooth CV $\theta$ and, using the identity

\begin{equation}\label{eqn:eqCurrent}
J_i\cdot \nabla f(x)=\frac{\pi(x)}{2}\left(\mathcal{L}f(x)-\mathcal{L}_{\pi}^{\dagger}f(x)\right),
\end{equation}
which follows from direct manipulations, we can write
\begin{equation}\label{eqn:JABgrad}
J_{AB}\cdot \nabla \theta(x)=
\frac{\pi(x)}{2}\left(q_-(x)\mathcal{L}[q_+\theta](x)-q_+(x)\mathcal{L}_{\pi}^{\dagger}[q_-\theta](x)\right).
\end{equation}
This formula is not useful computationally since it still contains a backward-in-time generator.  To compute statistics from data, we need to formulate their estimators as expectations against the stationary distribution since this (1) permits the use of the adjoint relation to clear away backward transition operators and (2) is consistent with our reweighting scheme.  To this end, we define the projected reactive current as
\begin{equation}
J_{AB}^\theta(s) = \int J_{AB}(x)\cdot \nabla \theta(x) \delta(\theta(x)-s)dx = 
\lim_{\lvert ds\rvert \rightarrow 0} \frac{1}{\lvert ds\rvert}\int_{\{\theta(x)\in ds\}}  J_{AB}(x)\cdot \nabla \theta(x)dx,
\end{equation}
where $ds$ is an infinitesimal region of CV space with $s\in ds$, and
$\{x:\theta(x)\in ds\}$ does not intersect $A\cup B.$
Using \eqref{eqn:JABgrad} and the fact that $\mathcal{L}q_+ = 0$ and $\mathcal{L}^\dagger_\pi q_- = 0$ on $(A\cup B)^\textrm{c}$, we have
\begin{align}
  J_{AB}^\theta(s) =   \lim_{\lvert ds\rvert \rightarrow 0} \frac{1}{\lvert ds\rvert}\int \1_{\{\theta(x) \in ds\}}
  \frac{\pi(x)}{2}&\left(q_-(x)\mathcal{L}[q_+\theta](x)-q_+(x)\mathcal{L}_{\pi}^{\dagger}[q_-\theta](x)\right) dx \nonumber\\
  =  \lim_{\lvert ds\rvert \rightarrow 0} \frac{1}{\lvert ds\rvert}\int \1_{\{\theta(x) \in ds\}}
  \frac{\pi(x)}{2}&\Big(q_-(x)\mathcal{L}[q_+\theta](x)- q_-(x)\mathcal{L}q_+(x) \theta(x) \nonumber\\
  &-q_+(x)\mathcal{L}_{\pi}^{\dagger}[q_-\theta](x)+q_+(x)\mathcal{L}_\pi^\dagger q_-(x) \theta(x)\Big) dx.
\end{align}
Writing this expression in terms of the transition operator and canceling terms, we find that
\begin{align}
 J_{AB}^\theta(s) &=  \lim_{t,\lvert ds\rvert \rightarrow 0} \frac{1}{2t\,\lvert ds\rvert}\int \1_{\{\theta(x) \in ds\}}
  \pi(x)\Big(q_-(x)\mathcal{T}^t[q_+\theta](x)- q_-(x)\mathcal{T}^tq_+(x) \theta(x)\notag \\
  &\hspace{2cm}
  -q_+(x)(\mathcal{T}^t)_{\pi}^{\dagger}[q_-\theta](x)
  +q_+(x)(\mathcal{T}^t)_\pi^\dagger q_-(x) \theta(x)\Big) dx\notag \\
&= \lim_{t,\lvert ds\rvert \rightarrow 0} \frac{1}{2t\,\lvert ds\rvert}\int 
  \pi(x)q_-(x)\Big( \1_{\{\theta(x) \in ds\}}\left(\mathcal{T}^t[q_+\theta](x)- \mathcal{T}^tq_+(x) \theta(x)\right)\notag \\
  &\hspace{2cm}
  +\left(\mathcal{T}^t[q_+\theta\, \1_{\{\theta \in ds\}}](x)-\mathcal{T}^t[q_+\1_{\{\theta \in ds\}}](x) \theta(x)\right)\Big) dx, \label{eqn::1}
\end{align}
where the second equality follows from the definition of the adjoint $(\mathcal{T}^t)^\dagger_\pi$.

Expression \eqref{eqn::1} for $J^\theta_{AB}(s)$ can be directly translated into an estimator for computing from short-trajectory data:
\begin{align}\label{eq:estimator_JABSup0}
    J^\theta_{AB}(s) \approx \frac{1}{2t\lvert ds \rvert}\sum_{i=1}^M  & q_+(X^{(i)}(t)) \left(\theta(X^{(i)}(t))-\theta(X^{(i)}(0))\right)\nonumber\\
    \times &q_-(X^{(i)}(0))\1_{\theta \in ds}(X^{(i)}(0))
     w(X^{(i)}(0))\nonumber\\
     + \frac{1}{2t\lvert ds \rvert}\sum_{i=1}^M & q_+(X^{(i)}(t)) \left(\theta(X^{(i)}(t))-\theta(X^{(i)}(0))\right)\nonumber\\
    \times & q_-(X^{(i)}(0))\1_{\theta \in ds}(X^{(i)}(t))
     w(X^{(i)}(0)).
\end{align}
Finally, without affecting the $t\rightarrow 0$ limit, we can stop our trajectories when they exit or enter $A\cup B$, yielding the estimator
\begin{align}\label{eq:estimator_JABSup}
    J^\theta_{AB}(s) \approx \frac{1}{2t\lvert ds \rvert}\sum_{i=1}^M & q_+(X^{(i)}(t\wedge T_{A\cup B})) \left(\theta(X^{(i)}(t\wedge T_{A\cup B}))-\theta(X^{(i)}(0))\right)\nonumber\\
    \times & q_-(X^{(i)}(0))\1_{\theta \in ds}(X^{(i)}(0))
     w(X^{(i)}(0))\nonumber\\
     + \frac{1}{2t\lvert ds \rvert}\sum_{i=1}^M & q_+(X^{(i)}(t)) \left(\theta(X^{(i)}(t))-\theta(X^{(i)}(S_{A\cup B}(t)))\right)\nonumber\\
    \times & q_-(X^{(i)}(S_{A\cup B}(t)))\1_{\theta \in ds}(X^{(i)}(t))
     w(X^{(i)}(0))
\end{align}
which, in our experience, outperformed \eqref{eq:estimator_JABSup0} for larger values of $t$.
Note that we could have canceled additional terms in  \eqref{eqn::1} to yield a more concise estimator.  However,  we found that the estimator \eqref{eq:estimator_JABSup} gave less noisy results.

\section{Reactive current on a CV space}
\label{secn:cvcurrent}
We now establish that our projected reactive current gives the flux over surfaces in CV space. We assume that our CVs are smooth and that, for some subset $C^\theta$ of CV space with smooth boundary, the set $C = \{x:\theta(x)\in C^{\theta}\}$ contains $A$ and does not intersect $B$.
We will establish that for such a subset,
\begin{equation}\label{eqn:def1}
\int_{\partial C^{\theta}}J^{\theta}_{AB}(s)\cdot n_{C^{\theta}} d\sigma_{C^{\theta}}=\int_{\partial C} J_{AB}\cdot n_C d\sigma_C.
\end{equation}
Here $n_{C^{\theta}}$ is the outward pointing normal vector to the boundary $\partial C^\theta$ of $C^\theta$, $n_C$ is the normal vector to the boundary $\partial C$ of $C$, $\sigma_{C^\theta}$ is the surface measure on $\partial C^\theta$ and, $\sigma_C$ is the surface measure on $\partial C$.  The significance of \eqref{eqn:def1} is that it shows that our definition of $J^\theta_{AB}$ preserves reactive flux across surfaces in the CV space so that statistics of reactive paths could, in principle, be computed directly from $J^\theta_{AB}$.

Let $f_\delta$ be a smooth function on CV space that is equal to 1 on $C^\theta$ and equal to 0 for $x$ a distance of more than $\delta$ from $C^\theta$.  Applying the divergence theorem and integrating by parts we find that
\begin{align}
\int_{\partial C^{\theta}}J^{\theta}_{AB}(s)\cdot n_{C^{\theta}} d\sigma_{C^{\theta}} &= \int_{C^\theta} \textrm{div}J^{\theta}_{AB}(s) ds \nonumber\\
&= \lim_{\delta\rightarrow 0} \int f_\delta(s)\, \textrm{div}J^{\theta}_{AB}(s) ds \nonumber\\
& = -\lim_{\delta\rightarrow 0} \int  J^{\theta}_{AB}(s)\cdot \nabla f_\delta(s) ds.
\end{align}
Inserting our definition of $J^\theta_{AB}$ we find that
\begin{align}
 \int_{\partial C^{\theta}}J^{\theta}_{AB}(s)\cdot n_{C^{\theta}} d\sigma_{C^{\theta}}
&= -\lim_{\delta\rightarrow 0} \sum_j \int \int J_{AB}(x)\cdot \nabla\theta_j(x)\delta(\theta(x)-s) \frac{\partial f_\delta(s)}{\partial s_j}dx ds  \nonumber\\
&= -\lim_{\delta\rightarrow 0} \sum_j \int J_{AB}(x)\cdot \nabla\theta_j(x) \frac{\partial f_\delta}{\partial s_j}(\theta(x))dx.
\end{align}
Using the chain rule the last expression can be rewritten as
\begin{equation}
\int_{\partial C^{\theta}}J^{\theta}_{AB}(s)\cdot n_{C^{\theta}} d\sigma_{C^{\theta}} = -\lim_{\delta\rightarrow 0}  \int J_{AB}(x)\cdot \nabla f_\delta(\theta(x)) dx.
\end{equation}
Integrating by parts, taking the $\delta\rightarrow 0$ limit, and applying the divergence theorem again yields \eqref{eqn:def1}.

\bibliographystyle{unsrt}
\bibliography{references,trp_cage_bib,dga_paper}

\clearpage


\beginsupplement

\begin{figure}[h!]
\includegraphics[scale=.4]{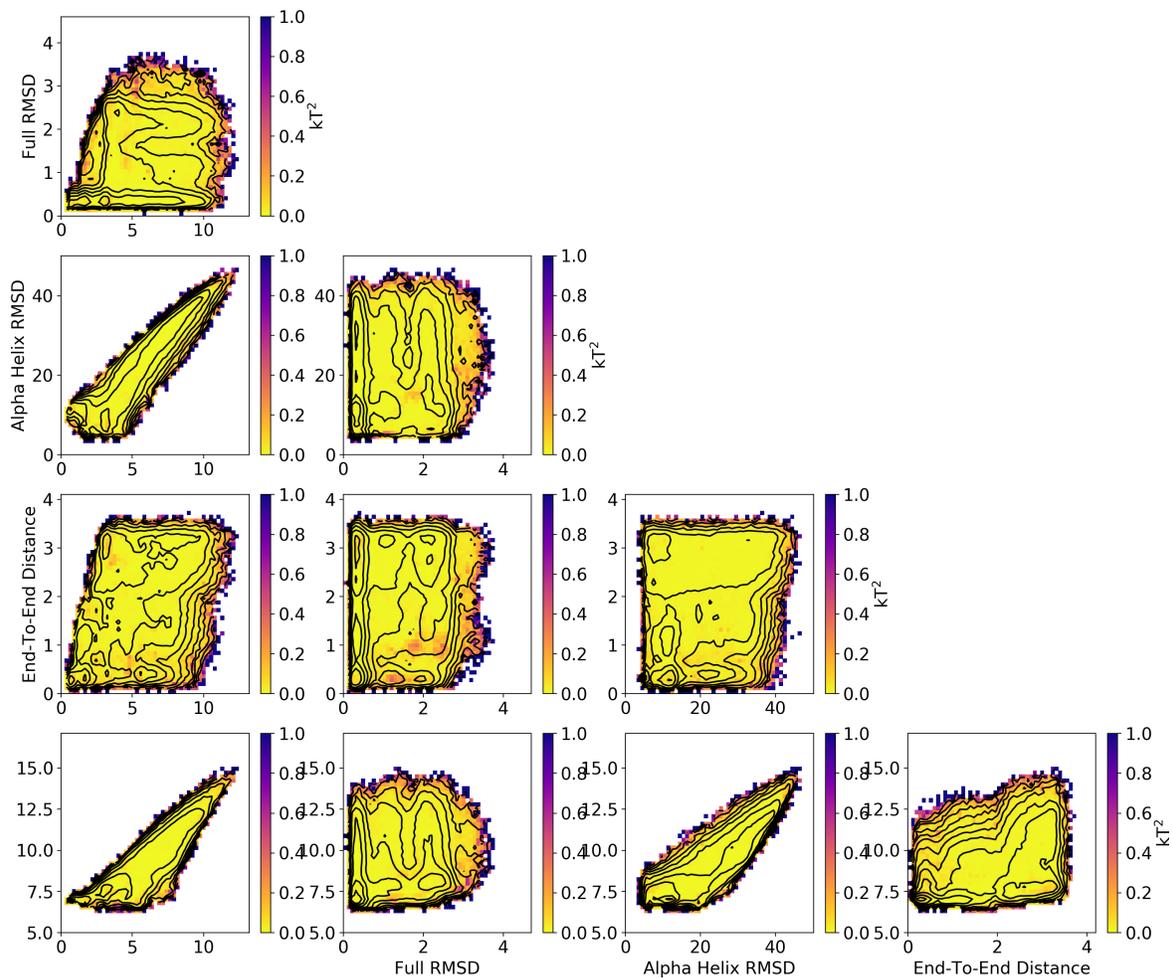}
\caption{
EMUS asymptotic variance for REUS PMFs.}
\label{fig:USAvars}
\end{figure}

\begin{figure}[h!]
\includegraphics[scale=.4]{figures_edited/UsampTriangle.png}
\caption{\label{fig:REUS}
REUS PMFs.}
\end{figure}

\begin{figure}[h!]
\includegraphics[scale=.4]{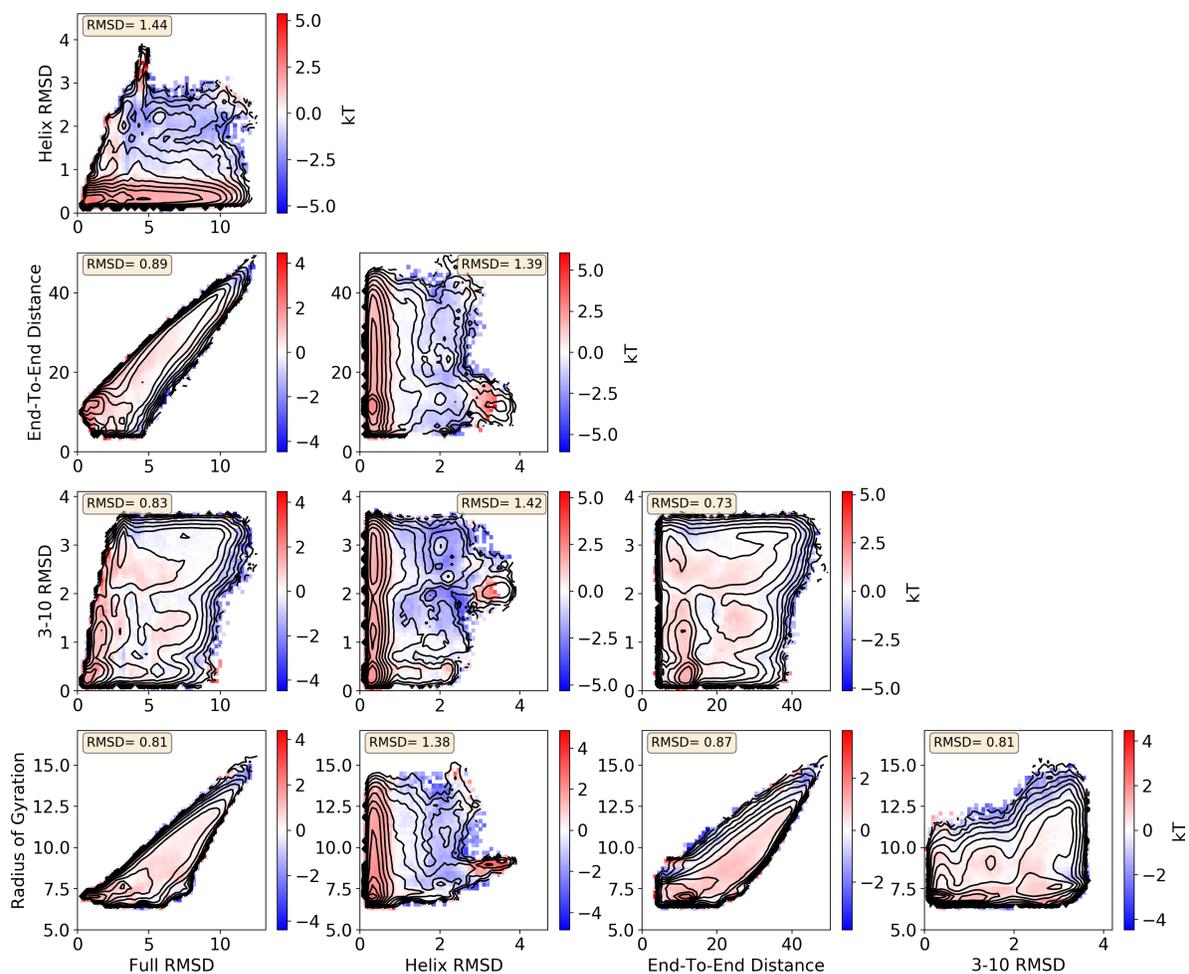}
\caption{\label{fig:USminusDGAInitialSampling}
Difference between DGA with the modified distance basis set without the $\alpha$ helix resampling and REUS.}
\end{figure}

\begin{figure}[h!]
\includegraphics[scale=.4]{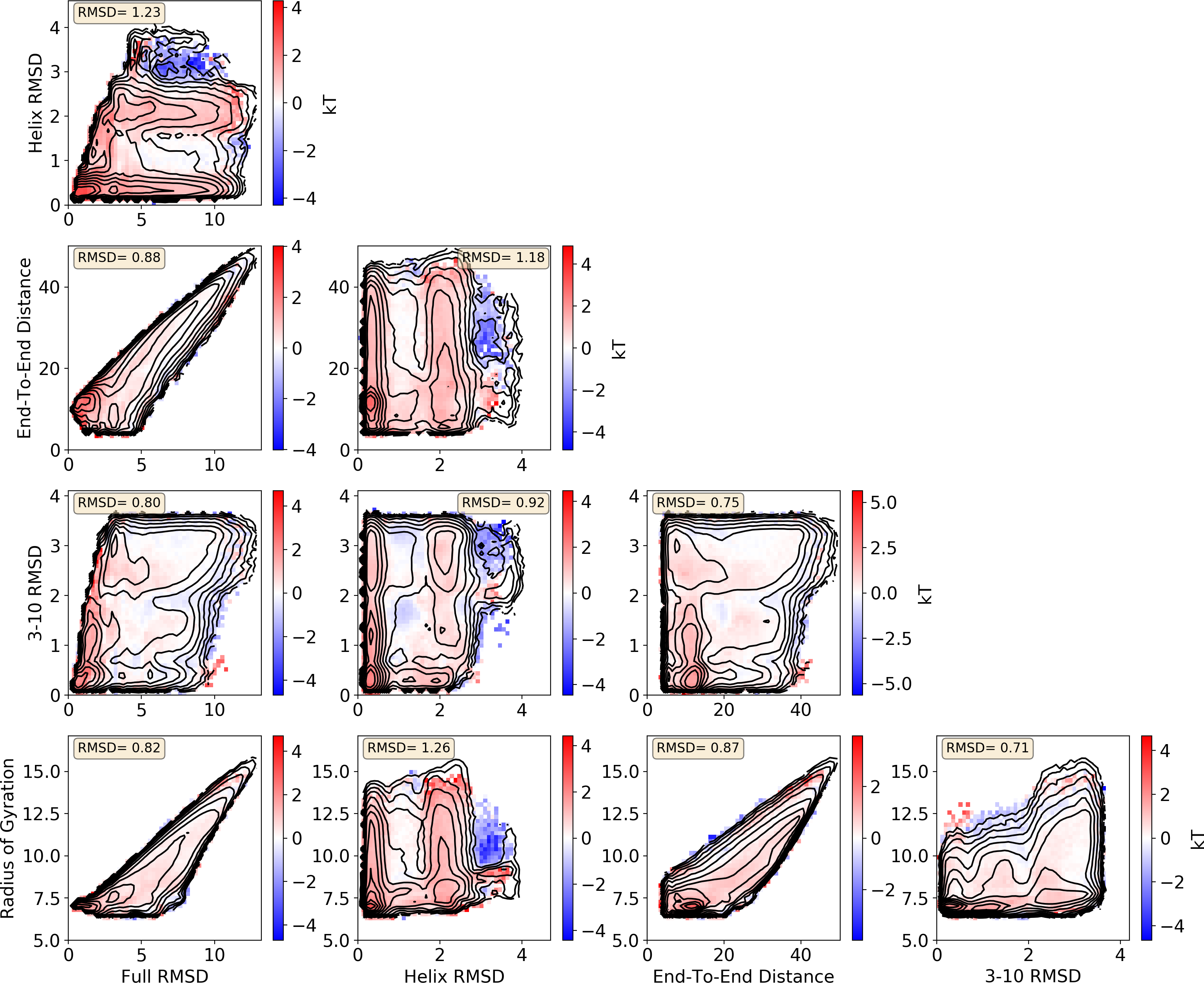}
\caption{\label{fig:USminusInd}
Difference between the PMF from DGA with the distance indicator basis set and the PMF from REUS.}
\end{figure}

\begin{figure}[h!]
\includegraphics[scale=.4]{figures_edited/Usamp_minus_TICA_Ind_DGA_Triangle_lag100.png}
\caption{\label{fig:USminusTICA}
Difference between the PMF from DGA with the TICA indicator basis set and the PMF from REUS.}
\end{figure}

\begin{figure}[h!]
\includegraphics[scale=.4]{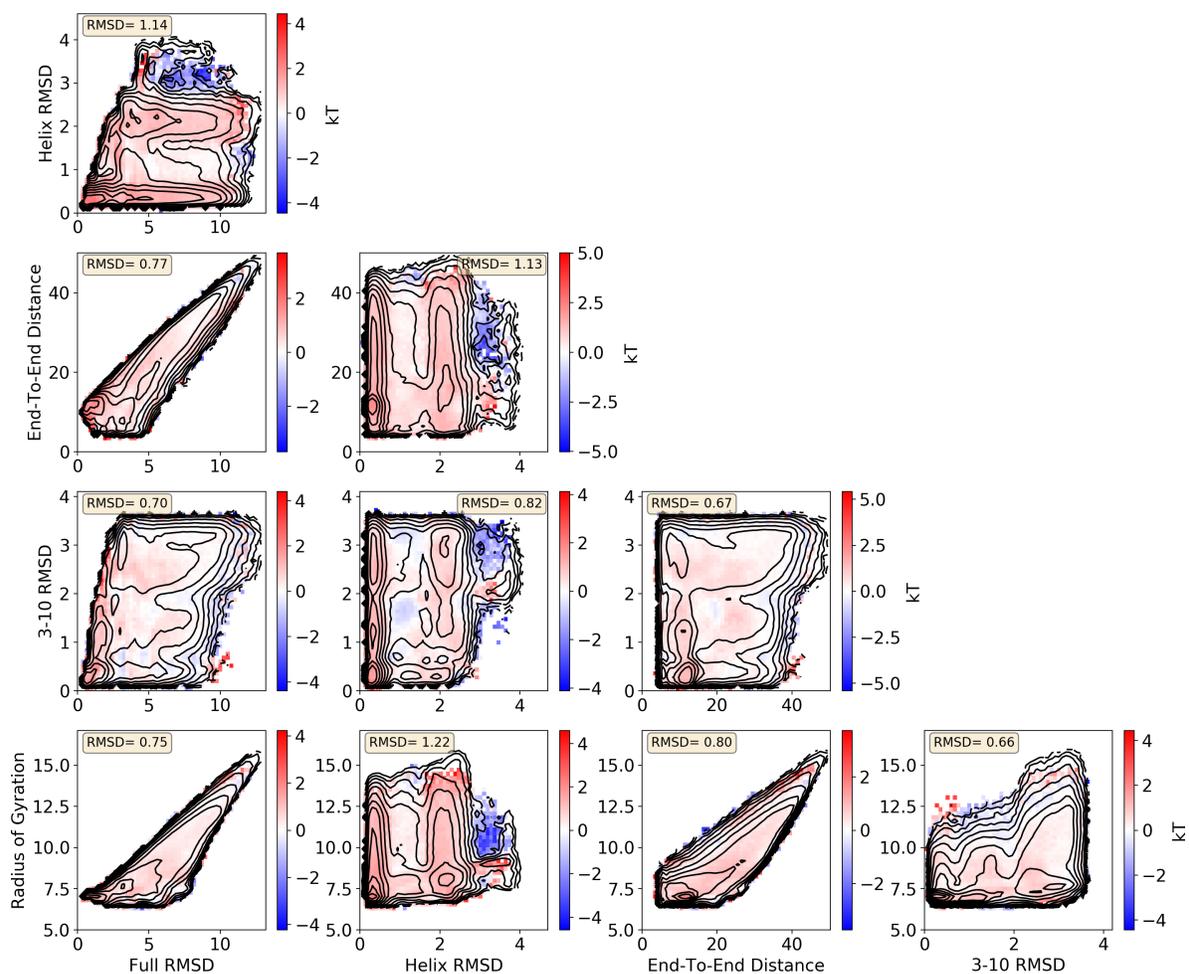}
\caption{\label{fig:USminusSmooth}
Difference between the PMF from DGA with the modified distance basis set and the PMF from REUS.}
\end{figure}

\begin{figure}[h!]
\includegraphics[scale=.9]{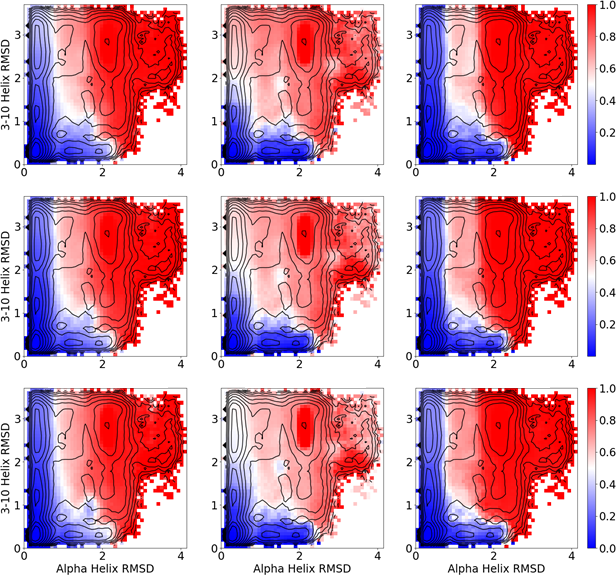}
\caption{\label{fig:backcomm}
DGA backward committors. Left, middle, and right columns are computed with the modified distance, distance indicator, and TICA indicator basis sets, respectively.  Top, middle, and bottom rows are computed with lag times of 0.5, 2.5, and 7.5 ns, respectively.}
\end{figure}

\makeatletter\@input{mainaux.tex}\makeatother